\pdfoutput=1
\documentclass[journal]{IEEEtran}

\usepackage[cmex10]{amsmath}
\usepackage{amssymb,amsfonts,mathrsfs,mathtools}
\usepackage{amsopn}

\usepackage{bm,bbm,dsfont}
\usepackage{cases}
\usepackage{accents}

\usepackage{algorithm}
\usepackage{algpseudocode}
\usepackage{algorithmicx}

\usepackage{array,xcolor}
\definecolor{revisionblue}{RGB}{0,90,180}

\usepackage{booktabs}
\usepackage{makecell}
\usepackage{multirow}
\usepackage{tabularx}
\usepackage[pdftex]{graphicx}
\DeclareGraphicsExtensions{.pdf,.jpeg,.png}

\usepackage{enumerate}
\usepackage{cite}
\usepackage{url,epstopdf}
\usepackage{xspace}
\usepackage[bookmarks=true,pageanchor,colorlinks,linkcolor=blue,anchorcolor=blue,citecolor=blue,urlcolor=blue,hyperfootnotes=false]{hyperref}

\newtheorem{theorem}{Theorem}

\newtheorem{proposition}{Proposition}
\newtheorem{definition}{Definition}
\newtheorem{assumption}{Assumption}
\newtheorem{lemma}{Lemma}
\newtheorem{remark}{Remark}

\long\def\old#1{}
\long\def\red#1{{#1}}

\long\def\blue#1{{#1}}
\IEEEoverridecommandlockouts

\def\BibTeX{{\rm B\kern-.05em{\sc i\kern-.025em b}\kern-.08em
		T\kern-.1667em\lower.7ex\hbox{E}\kern-.125emX}}

\begin{document}
\title{
Cone-Constrained Safe Policy Learning With Finite-Step Improvement Guarantees
}
\author{Liangliang~Hao,~Wei~Liu,~{and}~Yunjian~Xu%
\thanks{L. Hao and W. Liu are with the School of Control Science and Engineering, Shandong University, Jinan, China.
Y. Xu is with the Department of Mechanical Engineering and Automation, Chinese University of Hong Kong, Hong Kong SAR.
Emails: llhao@sdu.edu.cn, liuweikk@mail.sdu.edu.cn, yjxu@mae.cuhk.edu.hk.
{({\it Corresponding {author}: Liangliang Hao}).}
This work was supported in part by
National Natural Science Foundation of China (Grant No. 62503284), Natural Science Foundation of Shandong Province (Grant No. ZR2024QF242), and Key R\&D Program of Shandong Province (Grant No. 2026CXGC010305).
}}

\maketitle

\begin{abstract}
	One central challenge in safe reinforcement learning is improving task performance while {reducing infeasibility risk}, especially when long-term constraint consequences are difficult to represent and optimize.
	We propose the Constraint Aggregation Function (CAF), a learnable feasibility-oriented representation of long-horizon risk that satisfies {a} discounted Bellman recursion and admits a unique fixed point.
	The proposed CAF induces a risk advantage that measures the {reduction} in predicted unsafety relative to the current policy.
	Building on this representation, we develop Cone-Constrained Policy Optimization (CCPO), which uses the CAF-induced risk advantage to define a risk descent direction.
     {The proposed CCPO resolves} conflicts between task ascent and risk descent {through gradient projection}, selecting the final update from the relative interior of a constructed safety cone.
	Under standard smoothness assumptions {and a step-size bound}, this update guarantees simultaneous task improvement and risk reduction.
{Numerically, CCPO stays feasible with a 1.05\% mean optimality gap in emergency braking and attains low constraint violations on safety-critical benchmarks.}

\end{abstract}


\section{Introduction}
\label{intro}

Reinforcement learning (RL) has achieved strong performance in a wide range of sequential decision-making problems, including {continuous control~\cite{recht2019tour}}, EV charging scheduling~\cite{hao2022ev}, and real-world robot control{~\cite{brunke2022safe}}.
 However, in safety-critical applications such as EV charging operation and autonomous driving, a policy must not only maximize long-term return but also maintain safe behavior with near-zero violations. Achieving both safety and performance remains a central challenge for RL.

Safe RL is commonly formulated as a Constrained Markov Decision Process (CMDP)~\cite{altman1999cmdp}, where safety is encoded by a cost signal and handled by constraining the expected discounted cumulative cost{~\cite{li2023reinforcement, garcia2015comprehensive, borkar2005actor, wang2024safe}}. This formulation is simple and scalable, and has been widely used to reduce average violation cost{~\cite{wang2020vvc,wan2023adapsafe,shi2024alagev}}. However, expected cost primarily reflects safety in an aggregate sense. It does not directly describe whether the current state or action is pushing the system toward future infeasibility. As violations become rare, the resulting safety feedback also becomes sparse, making it difficult to learn a useful signal for persistent safety~\cite{ma2022joint}.

Once safety is difficult to represent, it also becomes difficult to optimize. In safe RL, one common line of methods handles the trade-off between return and safety through Lagrangian multiplier updates, which convert the constrained problem into an unconstrained objective and adjust the relative weight of reward and safety during learning~\cite{ray2019benchmarking,ding2021provably,stooke2020}. However, this optimization process can be unstable in practice~\cite{wu2024primaldual}, since the multiplier is sensitive to the changing balance between reward improvement and constraint satisfaction. More recently, a few works have instead examined the problem at the gradient level, seeking to reduce interference in policy updates by directly modifying or aggregating the reward and safety gradients{~\cite{kim2025conflictaverse, yao2024gradient, gu2024balance}}. Although these methods differ in how they optimize the policy, both rely on the safety objective to provide an informative signal about future safety.

These observations point to the need for a safety representation that captures how current decisions affect future safety and supports policy updates that improve safety and return together. {This paper addresses this need by introducing} a feasibility-oriented risk representation together with a projection-based policy update rule.

\subsection{Related Work}

Safe RL is commonly formulated as a constrained Markov decision process (CMDP), where safety is encoded through cost signals and optimized under expected cumulative-cost budgets~\cite{altman1999cmdp}. Policy optimization approaches include trust-region constrained optimization~\cite{achiam2017cpo}, Lagrangian and primal--dual methods{~\cite{ray2019benchmarking,stooke2020,paternain2023safe,dai2023augmented}}, first-order constrained policy optimization~\cite{zhang2020first}, reward--constraint switching~\cite{xu2021crpo}, penalized formulations~\cite{zhang2022p3o}, and constrained update projection~\cite{2022NIPS_CUP}. Although these methods differ in how the constraint is incorporated into policy optimization, they share an expectation-level view of safety based on cumulative costs. However, this formulation does not directly characterize whether safety can be maintained along future trajectories.

To address this limitation, studies have developed more direct representations of safety through probabilistic formulations{~\cite{chen2024tacprob,shen2024flipping,fan2025safety}}, safety certification{~\cite{ames2017cbfqp,chow2018lyapunov,cheng2019end,yang2023model}}, and feasible-set characterization~\cite{fisac2019bridging,yu2022reachability,ganai2023iterative}.
{Of particular relevance are methods that learn and characterize feasible sets.}
Hamilton--Jacobi (HJ) reachability{~\cite{mitchell2005hj}} has been connected with reinforcement learning through discounted dynamic-programming formulations, enabling value-based characterization of feasible sets and safety policies{~\cite{fisac2019bridging,akametalu2024discounted}}.
{RCRL~\cite{yu2022reachability} and~\cite{kokolakis2023reachability} learn reachability-based safety values and the associated safe sets from data, and RCRL further integrates them into constrained policy optimization.}
Feasible region iteration (FRI) further iteratively expands feasible regions toward the maximum feasible region~\cite{yang2023synthesizing}{, feasible policy iteration confines exploration to the feasible region of the current policy~\cite{zhang2025feasible}}, while RESPO learns a reachability estimation function for persistent state-wise safety~\cite{ganai2023iterative}.
{These methods use learned feasibility representations to characterize feasible sets or to constrain policy optimization, but seldom to derive an update direction that reduces future risk.}

Once a safety-side update direction is available, a further challenge is how to combine it with task ascent when the two directions conflict. Conventional scalarization or primal–dual methods regulate their relative influence through penalty or multiplier weights, but do not explicitly resolve the geometry of conflicting update directions. Gradient conflict has been widely studied in multi-task and multi-objective learning: PCGrad removes gradient components that oppose other objectives~\cite{yu2020gradient}, while CAGrad constructs conflict-averse aggregate directions~\cite{liu2021conflict}. Recent RL methods have also taken a gradient-based view, although they address different forms of conflict. GradS focuses on interactions among gradients induced by multiple safety constraints~\cite{yao2024gradient}, whereas CoMOGA studies gradient conflict among multiple objectives in constrained multi-objective RL~\cite{kim2025conflictaverse}. PCRPO more directly considers the conflict between reward and safety gradients, using {soft switching} and a slack mechanism to coordinate policy updates~\cite{gu2024balance}.
Overall, these methods address gradient conflict by coordinating competing gradients in multi-objective or multi-constraint settings.

{Closest to our work are FRI~\cite{yang2023synthesizing} and gradient-manipulation methods such as PCRPO~\cite{gu2024balance} and IMTL-G~\cite{liu2021imtl}.
FRI~\cite{yang2023synthesizing} represents feasibility by the constraint decay function, which is built from a binary violation indicator and used to label feasible regions.
{The proposed CAF extends the constraint decay function to a violation-graded, multi-constraint, action-conditioned signal, from which CCPO derives its risk-descent direction.}
PCRPO~\cite{gu2024balance} combines projected gradients with fixed equal weights and certifies only the summed objective, while IMTL-G~\cite{liu2021imtl} balances task gradients through equal projections in multi-task learning, without a safety signal and, to our knowledge, a joint step-size condition.
In contrast, the proposed CCPO combines the rectified gradients with norm-inverse weights. {We prove that, with exact gradients and a fixed step size, every conflicting update both increases the task return and decreases the policy-level CAF, and that the policy reaches near-minimal risk within finitely many iterations.}
}

\subsection{Summary of Results}

First, we propose the Constraint Aggregation Function (CAF), a feasibility-oriented representation of long-horizon safety {that generalizes the constraint decay function~\cite{yang2023synthesizing} to violation-graded, multi-constraint signals}. Unlike expectation-level costs in standard CMDP formulations~\cite{altman1999cmdp,ray2019benchmarking}, CAF is a bounded state--policy signal that tracks how unsafety propagates along future trajectories. It aggregates multiple constraint violations into an instantaneous unsafety surrogate and propagates this signal through a discounted Bellman recursion. The corresponding Bellman operator is a contraction, which guarantees a unique fixed point and provides step targets for critic learning. By further conditioning the recursion on actions, we obtain a CAF-induced risk action-value function and the corresponding risk advantage.

Second, we develop Cone-Constrained Policy Optimization (CCPO), a projection-based policy update method built on the CAF representation. CCPO turns the CAF-induced risk advantage into a risk-descent direction.
When the task-ascent and risk-descent directions conflict, CCPO removes their mutually opposing components through gradient projection. The resulting rectified directions span a safety cone, and the final policy update is selected from its relative interior using coefficients that balance their weighted magnitudes. This construction retains contributions from both task improvement and risk reduction while preventing direct opposition between the two update directions.

Third, we establish theoretical guarantees for the proposed cone-constrained update. We show that every direction in the relative interior of the safety cone forms an acute angle with both the task-ascent and risk-descent directions. Using smoothness-based quadratic bounds, we then derive a common step-size condition, independent of the conflict angle and the gradient-norm ratio, under which a single update simultaneously increases the task return and decreases the policy-level CAF objective.
We further prove that CCPO enters and remains in a prescribed risk region within an explicitly bounded number of iterations and then attains an {$\epsilon_{\mathrm{task}}$-stationary point of the task return and the risk objective} within this region.

Finally, we evaluate CCPO on an emergency-braking problem and four nonlinear safety-critical control benchmarks. In emergency braking, the learned CAF contour follows the reference feasible boundary over most of the state space. CCPO remains feasible on all evaluated states and achieves a 1.05\% mean raw gap to the convex optimum, whereas the constrained baselines exhibit larger optimality gaps or violate feasibility.
{Further experiments on the four nonlinear control benchmarks show that CCPO maintains low constraint violation rates while achieving competitive task returns.}

The remainder of this paper is organized as follows. Section~\ref{sec:prelim} introduces the safe RL formulation and the policy-dependent feasible set. Section~\ref{sec:method} develops CAF, constructs the cone-constrained policy update from the risk-descent direction, and presents the complete \textsc{CCPO} algorithm. Section~\ref{sec:theoretical_analysis} establishes alignment and finite-step guarantees for the projection-based policy update. Section~\ref{sec:exp} presents the numerical results, and Section~\ref{sec:conclusion} concludes the paper.

\section{Safe Reinforcement Learning and Feasible Set}
\label{sec:prelim}

\subsection{Preliminaries: Notation, Safety, and Problem Setup}
\label{subsec:preliminaries}

We study safe reinforcement learning in a deterministic Markov decision process (MDP) $\mathcal{M}=(\mathcal{X},\mathcal{U},f,r,\gamma,\rho_0)$. Here $\mathcal{X}$ and $\mathcal{U}$ are the state and action spaces, $f:\mathcal{X}\times\mathcal{U}\to\mathcal{X}$ is the dynamics, $r:\mathcal{X}\times\mathcal{U}\to\mathbb{R}$ is the reward, $\gamma\in(0,1)$ is the discount factor, and $\rho_0$ is the initial-state distribution. A stationary Markov policy $\pi(u|x)$ induces trajectories $\tau=(x_0,u_0,x_1,u_1,\ldots)$ with $x_0\sim\rho_0$, $u_t\sim\pi(\cdot|x_t)$, and $x_{t+1}=f(x_t,u_t)$. We denote its discounted state-visitation distribution by $d^\pi$.

The objective of standard RL is to maximize the expected discounted return
\begin{equation}
	J(\pi)\triangleq \mathbb{E}_{\tau\sim\pi}{\Bigl[}
	\sum_{t=0}^{\infty}\gamma^t r(x_t,u_t){\Bigr]}
	.
	\label{eq:standard_return}
\end{equation}
We model safety through $m$ constraint functions $\{g_i\}_{i=1}^m$, where $g_i(x)\le 0$ denotes satisfaction and $g_i(x)>0$ denotes violation.
For clarity we focus on state constraints. The same development applies to state--action constraints $g_i(x,u)\le 0$.
{The set of constraint-satisfying states is defined as follows.}

\begin{definition}[Safe set]
	\label{def:safe_set}
	The safe set is defined by
	\begin{equation}
		\mathcal{X}_{\mathrm{safe}}
		\triangleq
		\left\{\, x\in\mathcal{X}\ \big|\ g_i(x)\le 0,\ \forall i\in\{1,\ldots,m\}\,\right\}.
		\label{eq:safe_set}
	\end{equation}
\end{definition}

Definition~\ref{def:safe_set} describes whether the current state satisfies all constraints.
However, instantaneous safety does not guarantee future safety. A state can be safe now but still lead to violations later due to limited actuation or long-horizon closed-loop dynamics.
We therefore seek persistent safety under the policy-induced closed-loop evolution{~\cite{blanchini1999set}}.
This leads to a policy-dependent feasible set{.}

\begin{definition}[Policy-dependent feasible set]
	\label{def:feasible_set_pi}
	For a policy $\pi$, let the feasible set be
	\begin{equation}
		\mathcal{X}^{\pi}_{\mathrm{feas}}
		~\triangleq~
		\left\{x\in\mathcal{X}\ \middle|\
		\begin{aligned}
			&x_0=x,\ u_t\sim\pi(\cdot\mid x_t),\\
			&x_{t+1}=f(x_t,u_t),\\
			&\Pr_{\pi}\!\left(
			x_t\in\mathcal{X}_{\mathrm{safe}},\ \forall t\in\mathbb{N}
			\mid x_0=x\right)=1
		\end{aligned}
		\right\}
		\red{.}\label{eq:feasible_set_pi}
	\end{equation}
	If $\mathcal{X}^{\pi}_{\mathrm{feas}}\neq\emptyset$, we call $\pi$ feasible. Otherwise, it is infeasible.
\end{definition}

The two sets capture distinct notions of safety. The safe set records current constraint satisfaction, whereas the feasible set records whether that satisfaction persists under the closed-loop policy. Thus a state may lie in $\mathcal{X}_{\mathrm{safe}}$ but outside $\mathcal{X}^{\pi}_{\mathrm{feas}}$: it is safe at present, yet the policy may lead to a future violation. A policy is feasible when its policy-dependent feasible set is nonempty.

We seek a policy that maximizes the return in~\eqref{eq:standard_return} while promoting persistent safety in the sense of Definitions~\ref{def:safe_set}--\ref{def:feasible_set_pi}.
{%
Learning proceeds from rollouts with $x_0\sim\rho_0$, and the policy is updated so that trajectories from these initial states remain in $\mathcal{X}_{\mathrm{safe}}$.} In model-free settings, neither the dynamics nor the associated long-horizon safety boundary is known \emph{a priori}. This motivates a learned signal that summarizes how instantaneous violations propagate under the current policy and provides critic-like guidance for reducing future risk.

\section{CCPO: Cone-Constrained Policy Optimization}
\label{sec:method}
\begin{figure*}[t]
	\centering
	\includegraphics[width=0.9\textwidth]{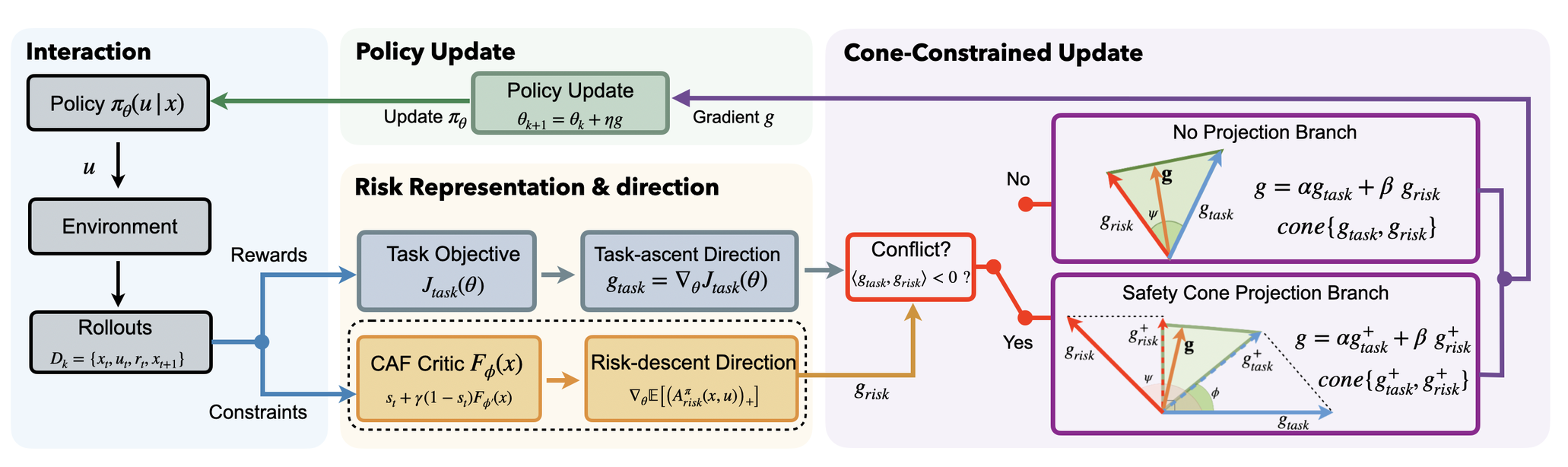}
	\caption{Overview of the CCPO framework. Rollouts collected under $\pi_\theta$ feed the task return $J_{\text{task}}(\theta)$
		and the CAF-induced risk advantage used to form the safety-side update. The task-ascent direction $g_{\text{task}}$ and risk-descent direction $g_{\text{risk}}$ define the policy update.}
	\label{fig:ccpo_overview}
\end{figure*}

This section constructs the Constraint Aggregation Function (CAF), derives from it a CAF-descent direction, resolves the conflict of that direction with task ascent through a cone-constrained update, and assembles the three into the CCPO algorithm. {Fig.~\ref{fig:ccpo_overview} summarizes the resulting pipeline.}

\subsection{Constraint Aggregation Function}
\label{sec:caf}
\label{sec:action_safety_q_adv}

We construct the Constraint Aggregation Function (CAF) as a bounded, policy-dependent representation of long-horizon infeasibility risk. Multiple constraint violations are first combined into a bounded instantaneous unsafety surrogate. Propagating this surrogate along policy rollouts yields the long-horizon representation. Conditioning the recursion on actions then yields a risk action-value and risk advantage that identify actions with elevated predicted risk.

Specifically, consider $m$ constraints $\{g_i(\cdot)\}_{i=1}^m$, where $g_i(x)\le 0$ indicates satisfaction and $g_i(x)>0$ indicates violation. The underlying violation measure is given by the positive-part operator $ g_i^+(x)=[g_i(x)]_+=\max\{g_i(x),0\}$,
which is zero exactly when the {$i$th} constraint is satisfied.
{For differentiable implementation, we use the softplus approximation, denoted by $\operatorname{sp}$,}
\begin{equation}
	\tilde g_i^{+}(x)\triangleq {\operatorname{sp}}(g_i(x))
	= \frac{1}{\kappa}\log\!\left(1+e^{\kappa g_i(x)}\right),
	\label{eq:smooth_pos}
\end{equation}
where $\kappa>0$ controls approximation sharpness. Thus, $\tilde g_i^{+}(x)\approx \max\{g_i(x),0\}$ while remaining smooth.
Softplus is therefore adopted for numerical implementation, whereas the exact zero-level characterization developed below is established for the underlying zero-preserving violation measure.

We aggregate the {constraint violations} with the log-sum-exp operator
\begin{equation}
	H(x) \triangleq \frac{1}{\vartheta}\log\!\left(1+\sum_{i=1}^{m}\exp\!\left(\vartheta\, g_i^{+}(x)\right)\right)
	-\frac{1}{\vartheta}\log(m+1),
	\label{eq:H}
\end{equation}
where $\vartheta>0$ controls sensitivity, with larger $\vartheta$ making $H(x)$ approach $\max_i g_i^{+}(x)$ (highlighting the most critical constraint) and smaller $\vartheta$ softly averaging across constraints.
The constant shift removes the nominal offset introduced by the log-sum-exp aggregation, and we always have $H(x)\ge 0$.
{Thus $H(x)=0$ if and only if $x\in\mathcal{X}_{\mathrm{safe}}$, and $H$ increases with each violation $g_i^+(x)$.}

Next, we map $H(x)$ to a bounded instantaneous unsafety indicator $s(x)\in[0,1)$:
\begin{equation}\label{eq:s_def}
	s(x) \triangleq 1-\exp\!\left(-\varpi H(x)\right),
\end{equation}
where $\varpi>0$ controls sensitivity.
This exponential squashing preserves ordering (larger $H$ implies larger unsafety) while keeping the signal bounded and numerically stable, which is convenient for long-horizon accumulation.
{In particular, $s(x)=0$ if and only if $x\in\mathcal{X}_{\mathrm{safe}}$. Compared with a binary violation indicator, $s(x)$ grows with the violation magnitude and thus yields a graded learning signal.}

Starting from this bounded instantaneous unsafety signal, we define CAF for the general dynamics $x_{t+1}=f(x_t,u_t)$ under a stochastic policy $\pi(u\mid x)$. A policy $\pi(u\mid x)$ induces a trajectory
\begin{equation}\label{eq:rollout_dynamics}
	u_t\sim\pi(\cdot\mid x_t),\qquad x_{t+1}=f(x_t,u_t),\qquad x_0=x.
\end{equation}

\begin{definition}[CAF]\label{def:caf}
	For a policy $\pi$ and discount factor $\gamma\in(0,1)$,
	the \textit{CAF} at state $x$ is defined as
	\begin{equation}\label{eq:caf}
		F^\pi(x)
		\mathrel{{\triangleq}} \mathbb{E}_{\tau\sim\pi}\!\left[\sum_{t=0}^{\infty}\gamma^t\, w_t\, s(x_t)\,\middle|\, x_0=x\right],
	\end{equation}
	where the \emph{survival weights} $\{w_t\}_{t\ge 0}$ are computed along the same rollout according to
	\begin{equation}\label{eq:survival_weights}
		w_0 \mathrel{{\triangleq}} 1,\qquad
		w_{t+1} \mathrel{{\triangleq}} w_t\bigl(1-s(x_t)\bigr),\quad t\in\mathbb{N}.
	\end{equation}
	Equivalently, the recursion admits the closed form
	\begin{equation}\label{eq:survival_weights_product}
		w_t = \prod_{j=0}^{t-1}\bigl(1-s(x_j)\bigr),\quad t\in\mathbb{N},
	\end{equation}
	with the convention $\prod_{j=0}^{-1}(\cdot)=1$.
\end{definition}
\noindent
The survival weights attenuate later contributions according to prior unsafety. {For the binary indicator $s=\mathbf 1[x\notin\mathcal X_{\mathrm{safe}}]$ (the limit $\varpi\to\infty$) and deterministic policies, $F^\pi(x)=\mathbb E[\gamma^{N^\pi(x)}]$ with $N^\pi(x)$ the first-violation time, i.e., the constraint decay function of~\cite{yang2023synthesizing}; as $\varpi\to0$, $F^\pi(x)/\varpi\to\mathbb E[\sum_{t\ge0}\gamma^tH(x_t)\mid x_0=x]$, a cost-type value.}
CAF therefore propagates risk along policy rollouts and admits the one-step recursion used for both analysis and learning. In practice, truncated rollouts and bootstrap targets approximate the infinite-horizon quantity.

We next establish that CAF is bounded and Bellman-consistent, making it suitable for fixed-point evaluation and learning.

\begin{proposition}[Nonnegativity and Boundedness]\label{prop:caf_bounded}
	For all $x \in \mathcal{X}$, $0 \le F^{\pi}(x) \le 1$.
\end{proposition}

\begin{IEEEproof}
	Since $s(x_t),w_t\in[0,1]$, we have $F^{\pi}(x)\ge 0$. Moreover,
	\begin{equation*}
		\begin{aligned}
			\sum_{t=0}^{T-1}\gamma^t w_t s(x_t)
			&\le \sum_{t=0}^{T-1} w_t s(x_t)
			= \sum_{t=0}^{T-1}(w_t-w_{t+1}) \\
			&= w_0 - w_T
			\le 1,
		\end{aligned}
	\end{equation*}
	for every $T\in\mathbb{N}$.
	Taking $T\to\infty$ and then expectation yields $F^{\pi}(x)\le 1$.
\end{IEEEproof}

\begin{proposition}[Bellman Consistency of CAF]\label{prop:bellman_caf}
	For every state $x\in\mathcal{X}$, the CAF satisfies the following Bellman-style recursion:
	\begin{equation}\label{eq:caf_bellman}
		F^\pi(x)
		=
	s(x)
	+
	\gamma\big(1-s(x)\big)\,
	\mathbb{E}_{u\sim\pi(\cdot\mid x)}
	\!\left[F^\pi\!\bigl(f(x,u)\bigr)\right].
\end{equation}
\end{proposition}

The proof of Proposition~\ref{prop:bellman_caf} is given in
Appendix~\ref{app:bellman_caf_proof}. This proposition provides
the one-step recursive characterization of CAF. The recursion underlies both the contraction analysis in Theorem~\ref{thm:contraction_mapping} and the practical bootstrap target used later for critic learning.

{%
We now establish the connection between the CAF $F^\pi$ and the policy-dependent feasible set $\mathcal{X}^{\pi}_{\mathrm{feas}}$ introduced in Definition~\ref{def:feasible_set_pi}.
This connection justifies interpreting $F^\pi$ as a \emph{learned feasibility indicator} rather than a generic cumulative-cost surrogate.

\begin{proposition}[CAF--feasibility equivalence]
	\label{prop:caf_feas_equiv}
	Let $f$ be deterministic and $\pi$ be a deterministic stationary policy. Then
	\begin{equation}
	F^\pi(x)=0 \ \Longleftrightarrow\ x\in\mathcal{X}^{\pi}_{\mathrm{feas}}.
	\label{eq:caf_feas_equiv}
	\end{equation}
\end{proposition}%
}
\begin{IEEEproof}
	($\Leftarrow$) If $x\in\mathcal{X}^{\pi}_{\mathrm{feas}}$, then $x_t\in\mathcal{X}_{\mathrm{safe}}$ for all $t\ge 0$ along the closed-loop trajectory, so $s(x_t)=0$ for all $t$ by construction. Definition~\ref{def:caf} then gives $F^\pi(x)=0$.
	($\Rightarrow$) Suppose $F^\pi(x)=0$ but $x\notin\mathcal{X}^{\pi}_{\mathrm{feas}}$. Then there exists $t^\star\ge 0$ with $s(x_{t^\star})>0$. Let $t^\star$ be the earliest such index; then $s(x_j)=0$ for $j<t^\star$, hence $w_{t^\star}=1$ and the $t^\star$-th term in \eqref{eq:caf} equals $\gamma^{t^\star}s(x_{t^\star})>0$. Since all terms are nonnegative, $F^\pi(x)>0$, a contradiction.
\end{IEEEproof}

{Under Proposition~\ref{prop:caf_feas_equiv}, $F^\pi=0$ characterizes feasibility, which means larger values indicate greater weighted unsafety.}
Proposition~\ref{prop:caf_feas_equiv} states that the zero-level set of $F^\pi$ exactly coincides with the feasible set. This equivalence further implies forward invariance of the CAF zero sublevel set.

{
\begin{proposition}[Forward invariance of the CAF zero sublevel set]\label{prop:caf_forward_inv}
Under the conditions of Proposition~\ref{prop:caf_feas_equiv},
the zero sublevel set
$\Omega_0^\pi \triangleq \{x\in\mathcal{X}:F^\pi(x)\le0\}$
is forward invariant under the closed-loop dynamics
$x_{t+1}=f(x_t,\pi(x_t))$.
Equivalently,
\begin{equation*}
	x_t\in\Omega_0^\pi
	\ \Longrightarrow\
	x_{t+1}\in\Omega_0^\pi.
\end{equation*}
\end{proposition}
\begin{IEEEproof}
    If $x_t\in\Omega_0^\pi$, then $F^\pi(x_t)\le0$.
    Nonnegativity in Proposition~\ref{prop:caf_bounded} gives $F^\pi(x_t)=0$.
    By Proposition~\ref{prop:caf_feas_equiv},
    $x_t\in\mathcal{X}_{\mathrm{feas}}^\pi$.
    Hence, the closed-loop trajectory remains feasible from
    $x_t$ onward, implying
    $x_{t+1}\in\mathcal{X}_{\mathrm{feas}}^\pi$.
    Applying Proposition~\ref{prop:caf_feas_equiv} again gives
    $F^\pi(x_{t+1})=0\le0$.
    Therefore, $\Omega_0^\pi$ is forward invariant.
\end{IEEEproof}
}

{Propositions~\ref{prop:caf_feas_equiv}--\ref{prop:caf_forward_inv} {establish} a feasibility interpretation of CAF: its zero sublevel set coincides with the policy-dependent feasible set and is forward invariant under the closed-loop dynamics.}
{Propositions~\ref{prop:caf_feas_equiv}--\ref{prop:caf_forward_inv} concern the exact violation measure $g_i^+$. With the softplus approximation, $\tilde g_i^+\le\log 2/\kappa$ on $\mathcal X_{\mathrm{safe}}$, so $F_\phi$ is small but not exactly zero on feasible states; in practice the zero level set is replaced by a small sublevel set of $F_\phi$, as used for the contour in Fig.~\ref{fig:ebc_combined}(b). The proof of Proposition~\ref{prop:caf_feas_equiv} also applies to stochastic $\pi$: $F^\pi(x)=0$ iff $\Pr^\pi(x_t\in\mathcal X_{\mathrm{safe}}\ \forall t)=1$.}
Section~\ref{sec:exp} illustrates this correspondence in the emergency-braking task.

Let $(\mathcal X,\Sigma_{\mathcal X})$ and $(\mathcal U,\Sigma_{\mathcal U})$ be measurable spaces. Assume that $s$ and $f$ are measurable and that $\pi(\cdot\mid x)$ is a Markov kernel. Let $\mathcal{B}$ be the Banach space of bounded measurable functions on $\mathcal{X}$ with norm $\|F\|_\infty\triangleq\sup_{x\in\mathcal{X}}|F(x)|$. {For fixed $\pi$ and $f$, define the CAF Bellman operator}
\begin{equation}
	\begin{aligned}
		(\mathcal{T}^\pi F)(x)
		&\triangleq s(x)+\gamma(1-s(x))\\
		&\quad\times\mathbb{E}_{u\sim\pi(\cdot\mid x)}
		\!\left[F\!\left(f(x,u)\right)\right].
	\end{aligned}
	\label{eq:Tpi_def}
\end{equation}

\begin{theorem}[Convergence of the CAF Bellman operator]
	\label{thm:contraction_mapping}
	For {$\gamma\in(0,1)$} and measurable $s:\mathcal{X}\to[0,1)$, $\mathcal{T}^\pi$ is a self-map and a $\gamma$-contraction on $(\mathcal{B},\|\cdot\|_\infty)$. Its unique fixed point is the CAF $F^\pi$, and $F^{(k+1)}=\mathcal{T}^\pi F^{(k)}$ satisfies $F^{(k)}\to F^\pi$ geometrically for every $F^{(0)}\in\mathcal{B}$.
\end{theorem}

Appendix~\ref{app:convergence_analysis_caf} provides the proof of Theorem~\ref{thm:contraction_mapping}. The contraction property of the exact CAF Bellman operator establishes a well-defined fixed-point target, thereby providing the theoretical foundation for learning CAF through neural Bellman regression.

We next expose the action dependence of the one-step CAF recursion by keeping the selected action explicit.
\begin{align}
Q_{\mathrm{risk}}^\pi(x,u)
&\triangleq s(x)+\gamma(1-s(x))F^\pi(f(x,u)),
\label{eq:Qrisk}\\
A_{\mathrm{risk}}^\pi(x,u)
&\triangleq F^\pi(x)-Q_{\mathrm{risk}}^\pi(x,u).
\label{eq:Arisk}
\end{align}
{By the Bellman consistency relation
$\mathbb{E}_{u\sim\pi}[Q_{\mathrm{risk}}^\pi(x,u)]=F^\pi(x)$,
$F^\pi(x)$ serves as the policy baseline, so
$A_{\mathrm{risk}}^\pi(x,u)>0$ ($<0$) indicates lower (higher) action-conditioned CAF.
Both quantities are obtained directly from the CAF critic and observed transitions, without an additional risk-$Q$ network
(Section~\ref{sec:algorithm}).}

\subsection{\texorpdfstring{{Cone-Constrained Policy Update}}{Cone-Constrained Policy Update}}
\label{sec:gradient_conflict}

{We now construct the cone-constrained policy update by integrating three components: the task-ascent direction, the CAF-induced risk-descent direction, and gradient projection. CAF provides long-horizon safety information at the action level, while the task gradient represents the return-improvement requirement. When these directions conflict, projection removes their mutually opposing components. The rectified directions then generate a cone that serves as the admissible set for policy optimization.}

{
Let $J_{\mathrm{task}}(\theta)$ denote the task-return objective in
\eqref{eq:standard_return}, with task-ascent direction
\begin{equation}
 g_{\mathrm{task}}(\theta)\triangleq\nabla_\theta J_{\mathrm{task}}(\theta).
 \label{eq:task_ascent_direction}
\end{equation}
{With the policy-level CAF objective $\mathcal R(\theta)\triangleq\mathbb E_{x_0\sim\rho_0}[F^{\pi_\theta}(x_0)]\in[0,1]$, the risk-descent direction is
\begin{equation}
	g_{\mathrm{risk}}(\theta)\triangleq-\nabla_\theta\mathcal R(\theta),
	\label{eq:task_risk_directions}
\end{equation}
which Section~\ref{sec:algorithm} estimates from $A_{\mathrm{risk}}^{\pi_\theta}$. Both directions are evaluated at the current iterate $\theta_k$.}
}

A central challenge is that the task-ascent and risk-descent directions may exhibit gradient conflict. In this case, directly combining them can introduce mutual cancellation, causing an update that improves one side to impair the other. We therefore first characterize their geometric relation before constructing the final policy update.

The directions $g_{\mathrm{task}}$ and $g_{\mathrm{risk}}$ conflict when
\begin{equation}
	\langle g_{\mathrm{task}},g_{\mathrm{risk}}\rangle<0,
	\label{eq:conflict_condition}
\end{equation}
so the two directions form an obtuse angle.

For nonzero $g_{\mathrm{task}}$ and $g_{\mathrm{risk}}$, the normalized conflict score
\begin{equation}
	\chi
	\triangleq
	\frac{\langle g_{\mathrm{task}},g_{\mathrm{risk}}\rangle}
	{\|g_{\mathrm{task}}\|_2\|g_{\mathrm{risk}}\|_2}
	\in[-1,1]
	\label{eq:conflict_score}
\end{equation}
separates angle from gradient scale. Values $\chi<0$ identify conflict, and values near $-1$ indicate nearly anti-parallel directions. This distinction matters because the projection geometry depends on angle, whereas the coefficients and learning rate also depend on gradient magnitude.

A fixed weighted combination of the two directions does not remove their opposing components. For example,
\begin{equation}
	g_{\lambda}
	=
	\lambda g_{\mathrm{task}}
	+(1-\lambda)g_{\mathrm{risk}},
	\qquad \lambda\in[0,1],
\end{equation}
can remain obtuse with one of its constituent directions. Let $c\triangleq-\langle g_{\mathrm{task}},g_{\mathrm{risk}}\rangle>0$ in the conflicting regime. Acute alignment of $g_\lambda$ with both constituent directions requires
\begin{equation}
	\frac{c}{\|g_{\mathrm{task}}\|_2^2+c}
	<\lambda<
	\frac{\|g_{\mathrm{risk}}\|_2^2}
	{\|g_{\mathrm{risk}}\|_2^2+c}.
	\label{eq:weighted_sum_interval}
\end{equation}
The admissible interval depends on the conflict angle and norm ratio, collapses for exactly anti-parallel directions, and shifts during training. A fixed $\lambda$ can therefore fall outside it.

When \eqref{eq:conflict_condition} holds, we remove from each direction the component that opposes the other by orthogonal projection:
\begin{align}
	g_{\mathrm{task}}^{+}
	&\triangleq
	g_{\mathrm{task}}-
	\frac{\langle g_{\mathrm{task}},g_{\mathrm{risk}}\rangle}
	{\|g_{\mathrm{risk}}\|_2^2}\,g_{\mathrm{risk}},
	\label{eq:g_task_plus}\\
	g_{\mathrm{risk}}^{+}
	&\triangleq
	g_{\mathrm{risk}}-
	\frac{\langle g_{\mathrm{risk}},g_{\mathrm{task}}\rangle}
	{\|g_{\mathrm{task}}\|_2^2}\,g_{\mathrm{task}}.
	\label{eq:g_risk_plus}
\end{align}
Thus, $g_{\mathrm{task}}^{+}$ removes the component of the task direction that opposes risk descent, and $g_{\mathrm{risk}}^{+}$ removes the component of the risk-descent direction that opposes task improvement.

The rectified directions satisfy the cross-orthogonality relations
\begin{equation}
	\left\langle g_{\mathrm{task}}^{+},g_{\mathrm{risk}}\right\rangle=0,
	\qquad
	\left\langle g_{\mathrm{risk}}^{+},g_{\mathrm{task}}\right\rangle=0.
	\label{eq:projection_cross_orthogonality}
\end{equation}
If $\psi=\angle(g_{\mathrm{task}},g_{\mathrm{risk}})$, then
\begin{equation}
	\|g_{\mathrm{task}}^{+}\|_2
	=\|g_{\mathrm{task}}\|_2\sin\psi,
	\qquad
	\|g_{\mathrm{risk}}^{+}\|_2
	=\|g_{\mathrm{risk}}\|_2\sin\psi.
	\label{eq:projected_direction_norms}
\end{equation}
Equation~\eqref{eq:projected_direction_norms} shows that each projected norm is scaled by $\sin\psi$. Both bases therefore shrink as $\psi\to\pi$, {so CCPO falls back to $g_{\mathrm{risk}}$ when $\sin\psi<\varepsilon_{\mathrm{ang}}$ for a prescribed $\varepsilon_{\mathrm{ang}}\in(0,1)$}. %

\begin{figure}[t]
	\centering
	\includegraphics[width=0.98\linewidth]{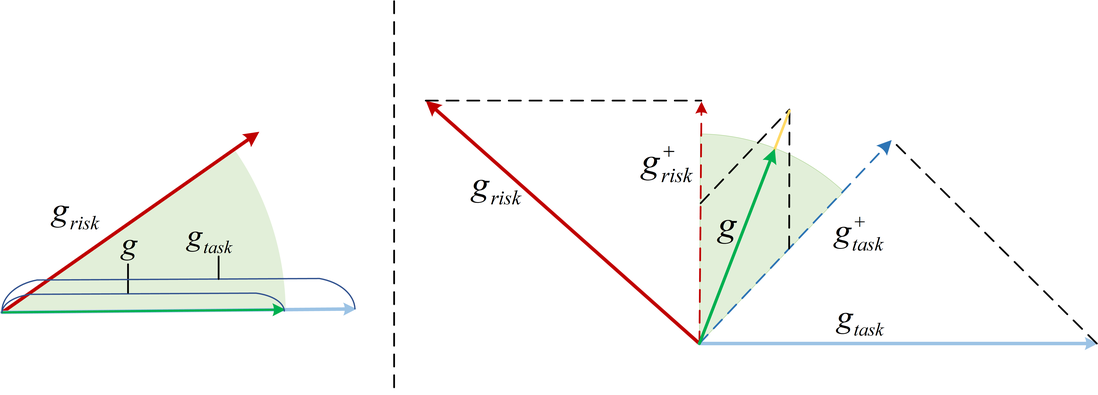}
	\caption{Geometry of the projection-based gradient update. When the task-ascent and risk-descent directions conflict, the mutually opposing components are removed and the policy update is selected from the relative interior of the safety cone spanned by the rectified directions.}
	\label{fig:gradmani}
\end{figure}

The rectified directions span the safety cone
\begin{equation}
	\begin{aligned}
	\mathcal{C}
	&\triangleq \mathrm{cone}\{g_{\mathrm{task}}^{+},g_{\mathrm{risk}}^{+}\}\\
	&=\{\alpha g_{\mathrm{task}}^{+}+\beta g_{\mathrm{risk}}^{+}:
	\alpha\ge 0,\ \beta\ge 0\}.
	\end{aligned}
	\label{eq:inner_direction}
\end{equation}
Because projection removes the mutually opposing components before the cone is formed, $\mathcal C$ is the set of conflict-compatible update directions. {Fig.~\ref{fig:gradmani} illustrates this geometry.}

CCPO constrains the final policy update to the relative interior of this cone:
\begin{equation}
	\operatorname{ri}(\mathcal{C})
	\triangleq
	\left\{
	\alpha g_{\mathrm{task}}^{+}+\beta g_{\mathrm{risk}}^{+}:
	\alpha>0,\ \beta>0
	\right\},
	\label{eq:inner_direction_neq}
\end{equation}
from which we select $g\in\operatorname{ri}(\mathcal{C})$ and update $\theta_{k+1}=\theta_k+\eta g$.

{%
To obtain a default relative-interior direction without letting gradient scale alone determine the update, set
\begin{equation}
	\alpha_k
	=\frac{\|g_{\mathrm{risk}}\|_2}{\|g_{\mathrm{task}}\|_2+\|g_{\mathrm{risk}}\|_2},
	\quad
	\beta_k
	=\frac{\|g_{\mathrm{task}}\|_2}{\|g_{\mathrm{task}}\|_2+\|g_{\mathrm{risk}}\|_2},
	\label{eq:cone_coefficient_choice}
\end{equation}
and select
\begin{equation}
	g_{\mathrm{cone}}
	=
	\alpha_k g_{\mathrm{task}}^{+}
	+
	\beta_k g_{\mathrm{risk}}^{+}.
	\label{eq:cone_direction_method}
\end{equation}
Strict positivity places $g_{\mathrm{cone}}$ in the relative interior rather than on a boundary ray, and the coefficients, being inversely proportional to the original gradient norms, equalize the magnitudes of the two weighted projected components.\footnote{{Unlike the fixed equal weights of PCRPO~\cite{gu2024balance}, the weights \eqref{eq:cone_coefficient_choice} make the direction of $g_{\mathrm{cone}}$ {under conflict} invariant to the scales of the CAF and return gradients.
Moreover, Theorem~\ref{thm:joint_monotonicity} certifies each objective separately with a step size independent of the conflict angle and norm ratio, whereas \cite{gu2024balance} {establishes monotone improvement of the summed objective with constant weights.}}} Section~\ref{sec:projection_monotonicity} {justifies this choice through} the finite-step monotonicity conditions.
}

When both directions are nonzero and
$\langle g_{\mathrm{task}},g_{\mathrm{risk}}\rangle\ge0$,
no opposing component needs to be removed. Every combination
$g_{\mathrm{task}}+\lambda_{\mathrm{risk}}g_{\mathrm{risk}}$
with $\lambda_{\mathrm{risk}}>0$ is positively aligned with both
directions. Projection is therefore activated only under
\eqref{eq:conflict_condition}. If one direction is zero, the
implementation uses the other. If both directions are zero, no
actor update is performed.

Under conflict, boundary rays are excluded because each is
orthogonal to one of the original directions, so every admissible
update receives a contribution from both rectified directions.

\begin{remark} %
	\label{rem:gating}
	By construction, $A_{\mathrm{risk}}^\pi(x,u)=\gamma(1-s(x))\big[{\mathbb{E}_{u'\sim\pi}[F^\pi(f(x,u'))]-F^\pi(f(x,u))}\big]$, so the advantage carries an implicit gate $\gamma(1-s(x))$ that attenuates its magnitude {as the violation grows}. This gating has two complementary effects: (i) it prevents the risk regularizer from dominating the update {once the constraints are violated} (where {immediate corrective power} is limited) {in the non-conflict branch (in the conflict branch the direction of $g_{\mathrm{cone}}$ is invariant to the scale of $g_{\mathrm{risk}}$, so the gate only reweights samples)}, and (ii) {on safe states the gate equals $\gamma$, so the learning signal is carried by the bracketed term, which is informative where a long-horizon distinction between actions persists}, i.e., approaching but not yet crossing the boundary. {The gate vanishes as $s\to1$, so the signal is most informative for avoiding, rather than recovering from, violations.}
\end{remark}

\subsection{CCPO Algorithm}
\label{sec:algorithm}

The resulting training pipeline combines CAF evaluation, risk-advantage estimation, risk-descent direction construction, conflict detection, and cone-constrained policy optimization. We learn a CAF critic $F_\phi$ as a state-value function that summarizes long-horizon unsafety under policy continuation. For each transition, the instantaneous unsafety $s_t \triangleq s(x_t)$ and the Bellman recursion define the bootstrap target
\begin{equation}
	\hat y_t
	=
	s_t
	+
	\gamma(1-s_t)F_{\bar\phi}(x_{t+1}),
	\label{eq:caf_target}
\end{equation}
where $F_{\bar\phi}$ is a slowly updated target network. The CAF critic minimizes
\begin{equation}
	\mathcal{L}_F(\phi)
	=
	\mathbb{E}_{t}
	\left[
	\bigl(F_\phi(x_t)-\hat y_t\bigr)^2
	\right].
	\label{eq:caf_loss}
\end{equation}

The state-value critic also provides an action-local risk-advantage estimate without requiring a separate risk-$Q$ network:
\begin{equation}
	\widehat A_{\mathrm{risk},t}
	=
	F_\phi(x_t)-\hat y_t.
	\label{eq:risk_adv_algo}
\end{equation}

For a rollout batch of size $N$, the task-ascent direction is estimated from the task advantage as
\begin{equation}
	\widehat g_{\mathrm{task}}
	=
	\frac{1}{N}
	\sum_{t=1}^{N}
	\widehat A_{\mathrm{task},t}
	\nabla_\theta \log \pi_\theta(u_t\mid x_t),
	\label{eq:sampled_task_direction}
\end{equation}
while the risk-descent direction is estimated from the CAF-induced risk advantage as
\begin{equation}
\widehat g_{\mathrm{risk}}
=
\frac{1}{N}
\sum_{t=1}^{N}
\widehat A_{\mathrm{risk},t}
\nabla_\theta \log \pi_\theta(u_t\mid x_t).
\label{eq:sampled_risk_direction}
\end{equation}

{The exact gradient of $\mathcal R$ weights each sample by $\gamma^t w_t$, which \eqref{eq:sampled_risk_direction} omits as a standard simplification also adopted in \eqref{eq:sampled_task_direction}.} Standard policy-gradient estimators may be used directly in
\eqref{eq:sampled_task_direction}--\eqref{eq:sampled_risk_direction}. In
practice, PPO-style clipped surrogates stabilize repeated minibatch
updates by limiting deviation from the rollout policy.
The projection rule of Section~\ref{sec:gradient_conflict} is then applied to the sampled directions, as summarized in Algorithm~\ref{alg:ccpo}.

{
Following exact-gradient analyses of gradient-manipulation methods such as PCRPO~\cite{gu2024balance}, PCGrad~\cite{yu2020gradient}, and CAGrad~\cite{liu2021conflict}, we connect the sampled actor updates to this geometry through the following vanishing-error consistency condition.
\begin{assumption}[Consistent actor-direction estimates]
	\label{assum:actor_consistency}
	As policy-evaluation and sampling errors vanish, the sampled actor directions satisfy
	\begin{equation*}
		\|\widehat g_{\mathrm{task}}-g_{\mathrm{task}}\|_2
		\xrightarrow{p}0,
		\qquad
		\|\widehat g_{\mathrm{risk}}-g_{\mathrm{risk}}\|_2
		\xrightarrow{p}0.
	\end{equation*}
\end{assumption}
Under Assumption~\ref{assum:actor_consistency}, the sampled update approaches the geometry defined by the exact directions in Section~\ref{sec:theoretical_analysis}.
}
The construction turns gradient conflict into a geometric direction-selection problem, whose finite-step convergence are established next.

\begin{algorithm}[t]
	\caption{\textsc{CCPO}}
	\label{alg:ccpo}
	\begin{algorithmic}[1]
		\State \textbf{Input:} $\theta_0,{\omega_0},\phi_0$, $\gamma,\eta,\lambda_{\mathrm{risk}},{\tau},\varepsilon_{\mathrm{proj}}{,\varepsilon_{\mathrm{ang}}}{,\kappa,\vartheta,\varpi}$
		\State Set target parameters $\bar\phi_0\leftarrow\phi_0$
		\For{$k=0,1,2,\dots$}
			\State Collect trajectories $\mathcal{D}_k=\{\tau_i\}$ using $\pi_{\theta_k}(u\mid x)$.
			\State Compute {returns and task advantages $\widehat A_{\mathrm{task},t}$}.
			\State Compute $s_t=s(x_t)$ and CAF targets $\hat y_t$ using \eqref{eq:caf_target}.
			\State Update ${V_\omega}$ using the task-value loss and update $F_\phi$ using \eqref{eq:caf_loss}.
			\State Compute $\widehat g_{\mathrm{task}}$ and $\widehat g_{\mathrm{risk}}$ using \eqref{eq:sampled_task_direction} and \eqref{eq:sampled_risk_direction}.
			\If{${\max}\{\|\widehat g_{\mathrm{task}}\|_2,
			\|\widehat g_{\mathrm{risk}}\|_2\}
			\le\varepsilon_{\mathrm{proj}}$}
				\State {No actor update; $g\leftarrow 0$.}
			\ElsIf{{$\min\{\|\widehat g_{\mathrm{task}}\|_2,
			\|\widehat g_{\mathrm{risk}}\|_2\}
			\le\varepsilon_{\mathrm{proj}}$}}
				\State {Set $g$ to whichever direction has norm $>\varepsilon_{\mathrm{proj}}$.}
			\ElsIf{{$\langle\widehat g_{\mathrm{task}},\widehat g_{\mathrm{risk}}\rangle<0$}}
				\State Rectify the sampled directions as in \eqref{eq:g_task_plus}--\eqref{eq:g_risk_plus}, obtaining $\widehat g_{\mathrm{task}}^{+}$ and $\widehat g_{\mathrm{risk}}^{+}$.
\State {If $\sin\widehat\psi<\varepsilon_{\mathrm{ang}}$, set $g\leftarrow\widehat g_{\mathrm{risk}}$, otherwise set $g\leftarrow g_{\mathrm{cone}}$ by \eqref{eq:cone_direction_method}.}
			\Else
				\State $g\leftarrow
				\widehat g_{\mathrm{task}}
				+\lambda_{\mathrm{risk}}
				\widehat g_{\mathrm{risk}}$
			\EndIf
			\State $\theta_{k+1}\leftarrow\theta_k+\eta g$
			\State $\bar\phi_{k+1}\leftarrow(1-{\tau})\bar\phi_k+{\tau}\phi_{k+1}$
		\EndFor
	\end{algorithmic}
\end{algorithm}

\section{Finite-Step Improvement Guarantees for Projected Policy Updates}
\label{sec:theoretical_analysis}

Section~\ref{sec:alignment_properties} establishes first-order alignment of projected directions, and Section~\ref{sec:projection_monotonicity} adds curvature control to obtain an admissible finite step.

\subsection{Alignment Properties of Projected Directions}
\label{sec:alignment_properties}

{The purpose of the projection is to construct one update direction with the correct first-order signs for task-return increase and decrease in $\mathcal R$. For $\theta^{+}=\theta+\eta g$, differentiability gives, as $\eta\downarrow0$,
\begin{align*}
	J_{\mathrm{task}}(\theta^{+})-J_{\mathrm{task}}(\theta)
	&=\eta\langle g_{\mathrm{task}},g\rangle+o(\eta),\\
	\mathcal R(\theta^{+})-\mathcal R(\theta)
	&=-\eta\langle g_{\mathrm{risk}},g\rangle+o(\eta).
\end{align*}
Thus, positive alignment with $g_{\mathrm{task}}$ and $g_{\mathrm{risk}}$ increases the task-return objective and decreases $\mathcal R$ for sufficiently small positive steps. Acute alignment is therefore the local geometric criterion for these two changes.}

Under conflict the projections \eqref{eq:g_task_plus}--\eqref{eq:g_risk_plus} satisfy the cross-orthogonality relations \eqref{eq:projection_cross_orthogonality}, so each projected basis vector is first-order neutral with respect to the other source direction. Non-collinearity ensures strictly positive alignment with its own source direction:
\begin{align}
	\left\langle g_{\mathrm{task}}^+,g_{\mathrm{task}}\right\rangle
	&=\|g_{\mathrm{task}}^+\|_2^2>0,\\
	\left\langle g_{\mathrm{risk}}^+,g_{\mathrm{risk}}\right\rangle
	&=\|g_{\mathrm{risk}}^+\|_2^2>0.
\end{align}
A positive combination of the projected bases therefore receives its task-improving component from $g_{\mathrm{task}}^+$ and its risk-descent component from $g_{\mathrm{risk}}^+$.

\begin{proposition}
	\label{prop:cone_acute_alignment}
	\emph{(Acute-angle alignment within the safety cone).}
	Let $g_{\mathrm{task}}, g_{\mathrm{risk}} \in {\mathbb{R}^p}$ be nonzero, non-collinear vectors satisfying the conflict condition $\langle g_{\mathrm{task}}, g_{\mathrm{risk}} \rangle < 0$.
	Let $g_{\mathrm{task}}^{+}$ and $g_{\mathrm{risk}}^{+}$ be the basis vectors defined by the projection operations in \eqref{eq:g_task_plus}--\eqref{eq:g_risk_plus}.
	For any $g\in\operatorname{ri}(\mathcal C)$,
	\begin{equation}
		\langle g,g_{\mathrm{task}}\rangle>0,
		\qquad
		\langle g,g_{\mathrm{risk}}\rangle>0.
	\end{equation}
	Thus, $g$ forms an acute angle with both the task-ascent direction and the risk-descent direction.
\end{proposition}
\begin{IEEEproof}
{The proof of Proposition~\ref{prop:cone_acute_alignment} is provided in Appendix~\ref{app:cone_acute_alignment}.}
\end{IEEEproof}

The projection also makes the source of each first-order improvement explicit.
For
\(
g=\alpha g_{\mathrm{task}}^++\beta g_{\mathrm{risk}}^+
\),
cross-orthogonality gives
\begin{align}
	\langle g,g_{\mathrm{task}}\rangle
	&=\alpha\|g_{\mathrm{task}}^+\|_2^2,
	\label{eq:main_task_alignment_decomposition}\\
	\langle g,g_{\mathrm{risk}}\rangle
	&=\beta\|g_{\mathrm{risk}}^+\|_2^2.
	\label{eq:main_risk_alignment_decomposition}
\end{align}
Projection therefore separates the two margins exactly, \(\alpha\) controlling the task side and \(\beta\) the risk side. This also fixes the role of the relative interior: a boundary coefficient sets one margin to zero, whereas positive coefficients make both strict.

The conflict angle determines the scale of these margins. Combining
\eqref{eq:main_task_alignment_decomposition}--\eqref{eq:main_risk_alignment_decomposition}
with \eqref{eq:projected_direction_norms} yields
\begin{align}
	\langle g,g_{\mathrm{task}}\rangle
	&=\alpha\|g_{\mathrm{task}}\|_2^2\sin^2\psi,\notag\\
	\langle g,g_{\mathrm{risk}}\rangle
	&=\beta\|g_{\mathrm{risk}}\|_2^2\sin^2\psi.
	\label{eq:main_alignment_angle_scaling}
\end{align}
The coefficients thus allocate the two margins and \(\sin^2\psi\) scales
both; the angle changes their magnitude but not the positive signs asserted
by Proposition~\ref{prop:cone_acute_alignment}.

\subsection{Task-Return Increase and Risk Decrease under Projection-Based Updates}
\label{sec:projection_monotonicity}

Finite-step guarantees require control of curvature in addition to favorable first-order terms. We derive admissible step-size bounds for task-return increase and CAF decrease, then fix the cone coefficients so that a single condition serves both.

To make the first-order intuition quantitative, consider an update $\theta_{k+1}=\theta_k+\eta g$. Its favorable linear terms are determined by the two alignment inner products $\langle g_{\mathrm{task}},g\rangle$ and $\langle g_{\mathrm{risk}},g\rangle$, while smoothness controls the accumulated gradient variation along the update.

\begin{assumption}[Smoothness on a compact parameter domain]
	\label{assum:smoothness}
	The policy parameter $\theta$ lies in a compact set $\Theta\subset\mathbb{R}^p$. The functions $J_{\mathrm{task}}(\theta)$ and $\mathcal R(\theta)$ are continuously differentiable on an open neighborhood of $\Theta$, and their gradients are Lipschitz on $\Theta$ {with constants} $L_{\mathrm{task}}>0$ and $L_{\mathrm{risk}}>0$, respectively. Every analyzed update segment $\{\theta_k+t\eta g:t\in[0,1]\}$ remains in $\Theta$.
\end{assumption}

Assumption~\ref{assum:smoothness} provides uniform quadratic control of the curvature terms along the update segment.
{The following standard lemma supplies the corresponding upper and lower models.}

\begin{lemma}\label{lem:descent_lemma}
	\emph{Smoothness quadratic model (descent/ascent lemma).}
{Let $\mathcal D\subseteq\mathbb{R}^p$, and let $h$ be differentiable on a neighborhood of $\mathcal D$ with an $L$-Lipschitz gradient on $\mathcal D$:}
	\begin{equation}\label{eq:Lsmooth_def}
		 {\|\nabla h(z)-\nabla h(z')\|_2\le L\|z-z'\|_2,
		\quad \forall z,z'\in\mathcal D.}
	\end{equation}
	 {For any $z,\Delta\in\mathbb{R}^p$ whose segment $\{z+t\Delta:t\in[0,1]\}$ lies in $\mathcal D$, we have}
	\begin{align}
		 {h(z+\Delta)}&\le {h(z)+\langle\nabla h(z),\Delta\rangle+\frac{L}{2}\|\Delta\|_2^2},
		\label{eq:descent_lemma}\\
		 {h(z+\Delta)}&\ge {h(z)+\langle\nabla h(z),\Delta\rangle-\frac{L}{2}\|\Delta\|_2^2}.
		\label{eq:ascent_lemma}
	\end{align}
\end{lemma}

Applying Lemma~\ref{lem:descent_lemma} to the task-return objective and $\mathcal R$ makes their first-order and curvature contributions explicit:
\begin{align}
	J_{\mathrm{task}}(\theta_k+\eta g)
	&\ge J_{\mathrm{task}}(\theta_k)
	+\eta\langle g_{\mathrm{task}},g\rangle
	-\frac{L_{\mathrm{task}}}{2}\eta^2\|g\|_2^2,
	\label{eq:task_smoothness_main}\\
	\mathcal R(\theta_k+\eta g)
	&\le \mathcal R(\theta_k)
	-\eta\langle g_{\mathrm{risk}},g\rangle
	+\frac{L_{\mathrm{risk}}}{2}\eta^2\|g\|_2^2.
	\label{eq:risk_smoothness_main}
\end{align}
For a direction with the positive task-side and risk-side alignments established in Proposition~\ref{prop:cone_acute_alignment}, the linear terms initially dominate their quadratic counterparts. The corresponding admissible upper bounds on the step size are
\begin{equation}
	\bar\eta_{\mathrm{task}}(g)
	\triangleq
	\frac{2\langle g_{\mathrm{task}},g\rangle}{L_{\mathrm{task}}\|g\|_2^2},
	\qquad
	\bar\eta_{\mathrm{risk}}(g)
	\triangleq
	\frac{2\langle g_{\mathrm{risk}},g\rangle}{L_{\mathrm{risk}}\|g\|_2^2}.
	\label{eq:direct_step_bounds_main}
\end{equation}
Their intersection preserves task-return increase and decrease in $\mathcal R$:
\begin{equation}
	0<\eta<\min\{\bar\eta_{\mathrm{task}}(g),\bar\eta_{\mathrm{risk}}(g)\}.
	\label{eq:joint_stepsize_main}
\end{equation}
These bounds quantify how far the local alignment can be followed before curvature overturns either desired change. Both depend on the selected cone direction.

The cone coefficients govern the two first-order margins together with the norm entering the quadratic remainder. A common scaling of $(\alpha,\beta)$ is absorbed into $\eta$, so only their ratio matters, and the normalized choice \eqref{eq:cone_coefficient_choice} balances the weighted magnitudes of the two projected directions. Because both projected norms carry the same $\sin\psi$ factor, it gives $\alpha_k\|g_{\mathrm{task}}^+\|_2=\beta_k\|g_{\mathrm{risk}}^+\|_2$: the two contributions enter $g_{\mathrm{cone}}$ with equal magnitude, so unequal gradient scales alone cannot determine the direction.

For this coefficient choice, Appendix~\ref{app:stepsize_bounds} shows that the two admissible upper bounds exceed $1/(2L_{\mathrm{task}})$ and $1/(2L_{\mathrm{risk}})$, respectively. Therefore,
\begin{equation}
	0<\eta<
	\min\!\left\{
	\frac{1}{2L_{\mathrm{task}}},
	\frac{1}{2L_{\mathrm{risk}}}
	\right\}
	\label{eq:conservative_joint_stepsize}
\end{equation}
is sufficient for task-return increase and decrease in $\mathcal R$. Hence, \eqref{eq:conservative_joint_stepsize} provides a common admissible step size for the projected update.
Combining projection, smoothness, and the coefficient choice gives the following guarantee, which holds for any conflict angle and gradient-norm ratio.

\begin{theorem}[Joint Task Improvement and Risk Reduction]
	\label{thm:joint_monotonicity}
	{Suppose Assumption~\ref{assum:smoothness} holds.} {Let $g_{\mathrm{task}}$ and $g_{\mathrm{risk}}$ be nonzero and non-collinear, and suppose $\langle g_{\mathrm{task}},g_{\mathrm{risk}}\rangle<0$.} Construct $g_{\mathrm{cone}}$ from \eqref{eq:g_task_plus}--\eqref{eq:g_risk_plus} using the coefficients in \eqref{eq:cone_coefficient_choice}, and set $\theta_{k+1}=\theta_k+\eta g_{\mathrm{cone}}$. If the update segment remains in $\Theta$ and \eqref{eq:conservative_joint_stepsize} holds, then
	\begin{equation}
		J_{\mathrm{task}}(\theta_{k+1})>J_{\mathrm{task}}(\theta_k),
		\qquad
		\mathcal R(\theta_{k+1})<\mathcal R(\theta_k).
	\end{equation}
\end{theorem}
\begin{IEEEproof}
	Proposition~\ref{prop:cone_acute_alignment} gives positive task-side and risk-side alignment terms. Appendix~\ref{app:stepsize_bounds} shows that the chosen step size satisfies both admissibility conditions. Substitution into \eqref{eq:task_smoothness_main}--\eqref{eq:risk_smoothness_main} yields the result.
\end{IEEEproof}

{Theorem~\ref{thm:joint_monotonicity} shows that, despite gradient conflict, a single projected step increases the task return and decreases the policy-level CAF, with a step size \eqref{eq:conservative_joint_stepsize} independent of the conflict angle and the gradient-norm ratio. By \eqref{eq:main_alignment_angle_scaling}, the angle affects only the size of the improvement, not its sign.}

{
{The alignment and monotonicity results above extend to a finite-iteration guarantee under the following condition.}
{\begin{assumption}[Uniform gradient domination]
\label{assum:gradient_domination}
There exists $c_{\mathrm{risk}}>0$ such that, for all $\theta\in\Theta$,
\begin{equation}
\|g_{\mathrm{risk}}(\theta)\|_2
\ge
c_{\mathrm{risk}}
\bigl[\mathcal R(\theta)-\mathcal R^\star_\Theta\bigr],
\label{eq:risk_gradient_domination}
\end{equation}
where $\mathcal R^\star_\Theta\triangleq\min_{\theta\in\Theta}\mathcal R(\theta)$.
\end{assumption}}
{For finite tabular softmax policies, \cite[Lemma~8]{mei2020global} yields a bound of this form with a policy-dependent constant, which is bounded away from zero on the compact set $\Theta$.
}
{\begin{theorem}[Finite-Step Improvement Guarantees]
\label{thm:finite_step}
Suppose Assumptions~\ref{assum:smoothness} and~\ref{assum:gradient_domination} hold, {all iterates remain in $\Theta$,} $|r|\le r_{\max}$, and Algorithm~\ref{alg:ccpo} uses exact directions with a sufficiently small constant step size $\eta$. Fix tolerances $\epsilon_{\mathrm{risk}}\in(0,1]$ and $\epsilon_{\mathrm{task}}>0$, and let $\varepsilon_{\mathrm{proj}}<\min\{\epsilon_{\mathrm{task}},c_{\mathrm{risk}}\epsilon_{\mathrm{risk}}/2\}$ and $\varepsilon_{\mathrm{ang}}\le\epsilon_{\mathrm{task}}/G_{\mathrm{task}}$, where $G_{\mathrm{task}}\triangleq\max_{\Theta}\|g_{\mathrm{task}}\|_2$. Then there exists a finite $K_{\mathrm{in}}$ such that
\begin{equation}
\mathcal R(\theta_k)
\le
\mathcal R^\star_\Theta+\epsilon_{\mathrm{risk}},
\qquad k\ge K_{\mathrm{in}},
\label{eq:finite_risk_retention}
\end{equation}
and a finite $k^\star\ge K_{\mathrm{in}}$, bounded explicitly in \eqref{eq:finite_step_bound}, such that
\begin{equation}
\min_{\mu\ge0}
\|g_{\mathrm{task}}(\theta_{k^\star})
+\mu g_{\mathrm{risk}}(\theta_{k^\star})\|_2
\le
\epsilon_{\mathrm{task}}.
\label{eq:finite_task_stationarity}
\end{equation}
\end{theorem}}
{Theorem~\ref{thm:finite_step} shows that CCPO enters the prescribed risk region in finitely many iterations and then attains an $\epsilon_{\mathrm{task}}$ stationary point within that region. Its proof is given in Appendix~\ref{app:finite_step_convergence}.}
}

\section{Numerical Results}
\label{sec:exp}

Section~\ref{sec:emergency_braking} examines feasible-boundary recovery and policy quality on an analytically interpretable emergency-braking task, {and} Section~\ref{sec:safety_control_benchmarks} evaluates the safety--performance trade-off on four nonlinear control benchmarks.

\subsection{Emergency Braking Control}
\label{sec:emergency_braking}

\begin{figure}[t]
	\centering
	\begin{minipage}[t]{0.48\linewidth}
		\centering
		\includegraphics[width=\linewidth]{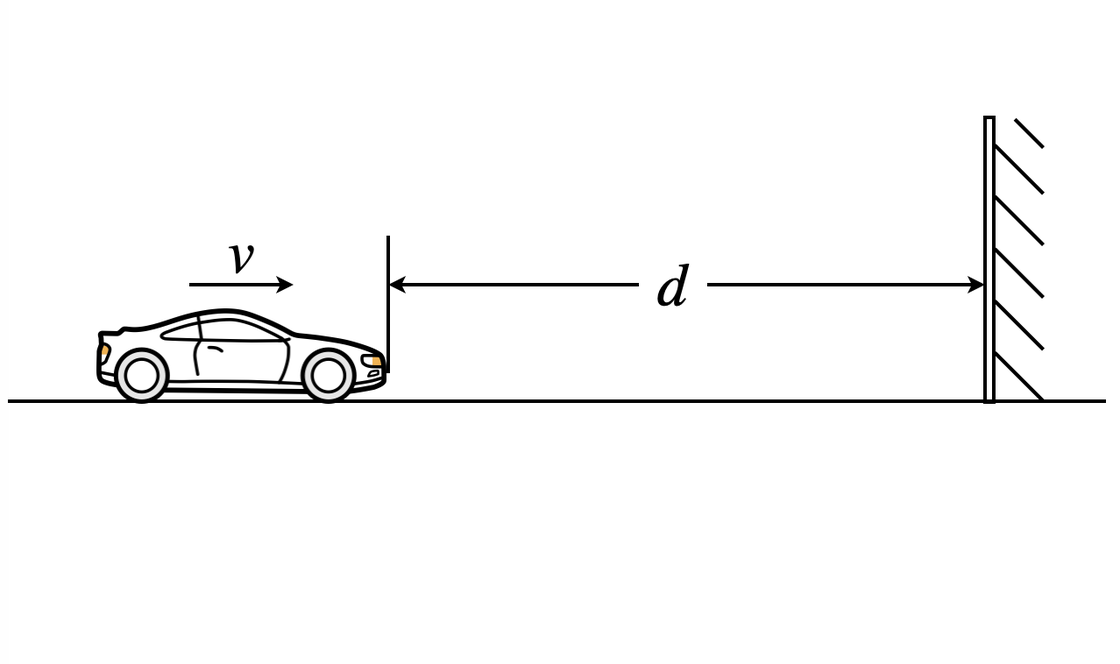}
		\vspace{2pt}
		{\footnotesize (a) Emergency braking setup.}
	\end{minipage}
	\hfill
	\begin{minipage}[t]{0.48\linewidth}
		\centering
		\includegraphics[width=\linewidth]{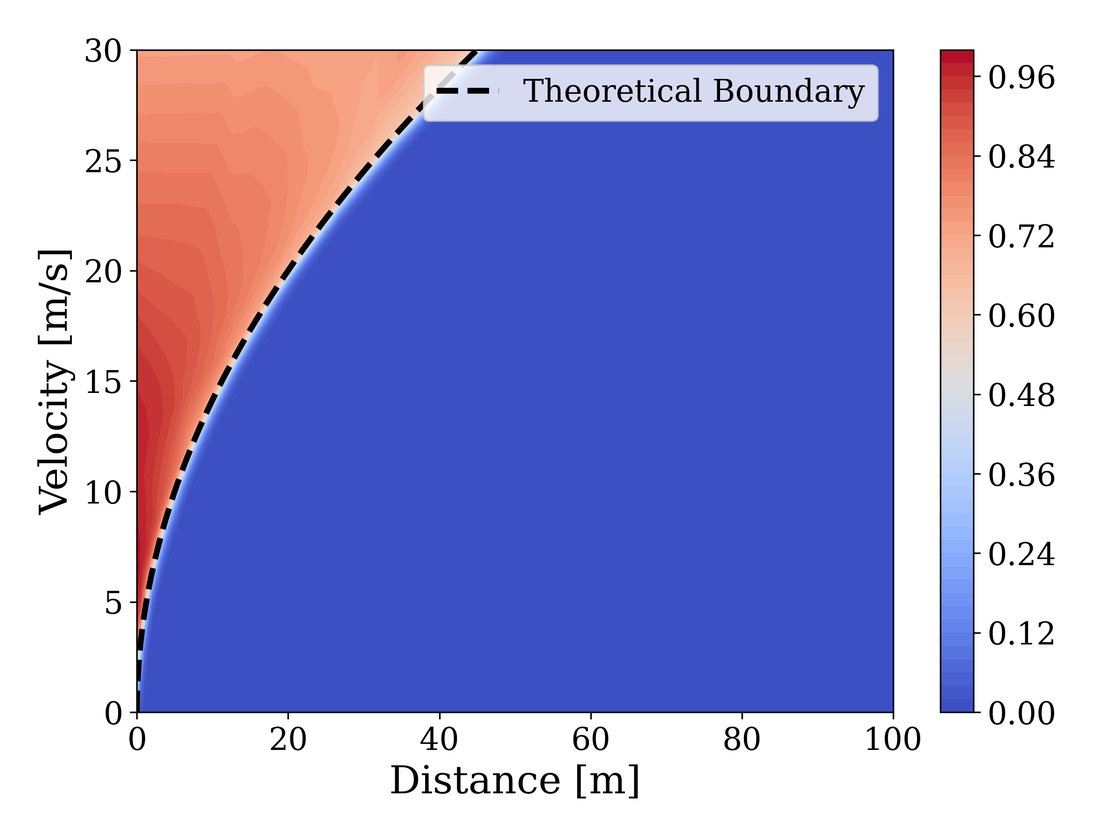}

		\vspace{2pt}
		{\footnotesize (b) CAF boundary contour.}
	\end{minipage}

		\caption{Emergency-braking setup and comparison between the CAF contour and reference feasible boundary.}
	\label{fig:ebc_combined}
\end{figure}

We consider a vehicle approaching a static obstacle, as illustrated in Fig.~\ref{fig:ebc_combined}(a). The state is $x_t=[d_t,v_t]^\top$, where {$d_t$} is the distance to the obstacle and $v_t\ge 0$ is the longitudinal velocity. The control input is the acceleration $u_t=a_t\in[a_{\min},0]$, with $a_{\min}<0$ denoting maximum braking. The dynamics follow a discrete-time longitudinal model
\begin{equation}
	\begin{bmatrix}
		d_{t+1}\\
		v_{t+1}
	\end{bmatrix}
	=
	\begin{bmatrix}
		1 & -\Delta t\\
		0 & 1
	\end{bmatrix}
	\begin{bmatrix}
		d_t\\
		v_t
	\end{bmatrix}
	+
	\begin{bmatrix}
		0\\
		\Delta t
	\end{bmatrix}a_t,
	\label{eq:ebc_dynamics}
\end{equation}
where $\Delta t$ is the sampling time. {The finite-horizon task cost penalizes braking effort, acceleration changes, running speed, and terminal speed.}
Safety is defined by the state constraint
\begin{equation}
	 g(d_t, v_t) \triangleq -d_t \le 0
	 \quad (\text{i.e., } d_t \ge 0){.}
	\label{eq:ebc_constraint}
\end{equation}
Thus, $d_t=0$ defines the safety boundary, and $d_t<0$ indicates collision.

Dense-grid rollouts under maximum braking provide a monotone reference boundary in the $(d,v)$ plane, where higher velocities require longer stopping distances. {This boundary delimits the largest feasible set, which contains $\mathcal{X}^{\pi}_{\mathrm{feas}}$ for every policy.} The CAF contour closely follows this reference over most of the state space, with larger deviations confined to sparsely explored regions (Fig.~\ref{fig:ebc_combined}(b)).

\begin{figure}[t]
	\centering
	\includegraphics[width=\columnwidth]{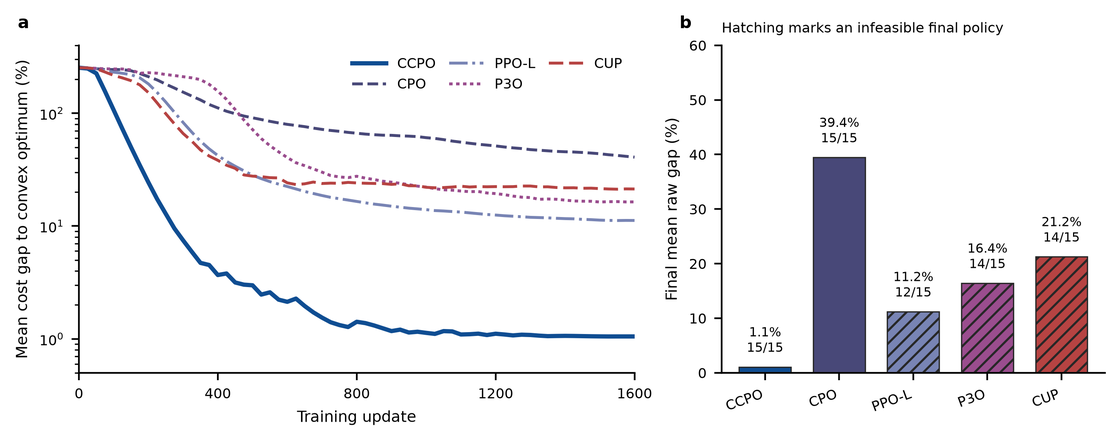}
	\caption{{CCPO and baseline performance. (a) Mean gap over 1,600 updates. (b) Final mean gap and physical-feasibility count on 15 initial states.}}
	\label{fig:ebc_safe_rl_comparison}
\end{figure}

\begin{figure}[t]
	\centering
	\includegraphics[width=\columnwidth]{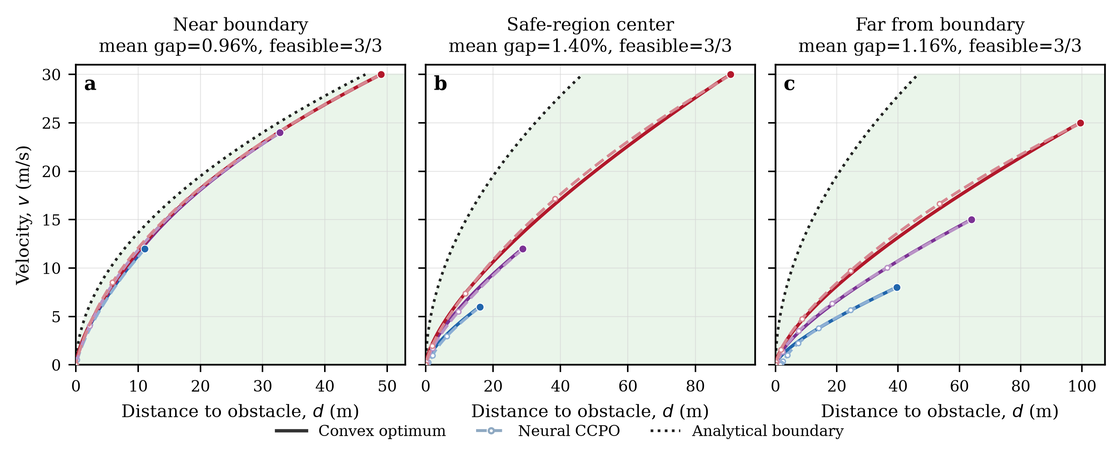}
	\caption{{CCPO and convex-optimal trajectories for representative initial states. The dotted curve marks the feasible boundary, and shading denotes the feasible region.
	}}
	\label{fig:ebc_trajectory_comparison}
\end{figure}

{We compare CCPO with CPO, PPO-Lag, P3O, and CUP~\cite{achiam2017cpo,ray2019benchmarking,zhang2022p3o,2022NIPS_CUP} under a common horizon and update budget, and evaluate each policy from the same 15 initial states. The analytical stopping boundary and the per-state convex optimum serve only as evaluation references, and performance is measured by the relative gap %
between the policy cost %
and the optimal cost %
value (via dynamic programming). As summarized in Fig.~\ref{fig:ebc_safe_rl_comparison}, at update 1,600 CCPO attains full physical feasibility with a 1.05\% mean raw gap. CPO is also feasible but has a 39.44\% mean gap, whereas PPO-Lag, P3O, and CUP violate the physical stopping condition for at least one initial state. An additional trajectory evaluation yields full feasibility, with mean and worst gaps of 1.42\% and 4.90\%, respectively. As shown in Fig.~\ref{fig:ebc_trajectory_comparison}, CCPO almost coincides with the convex optimum near the feasible boundary, where safety is most critical, and deviates only modestly near the center of the feasible region.}

\begin{figure*}[htbp]
	\centering
	\includegraphics[width=0.98\textwidth]{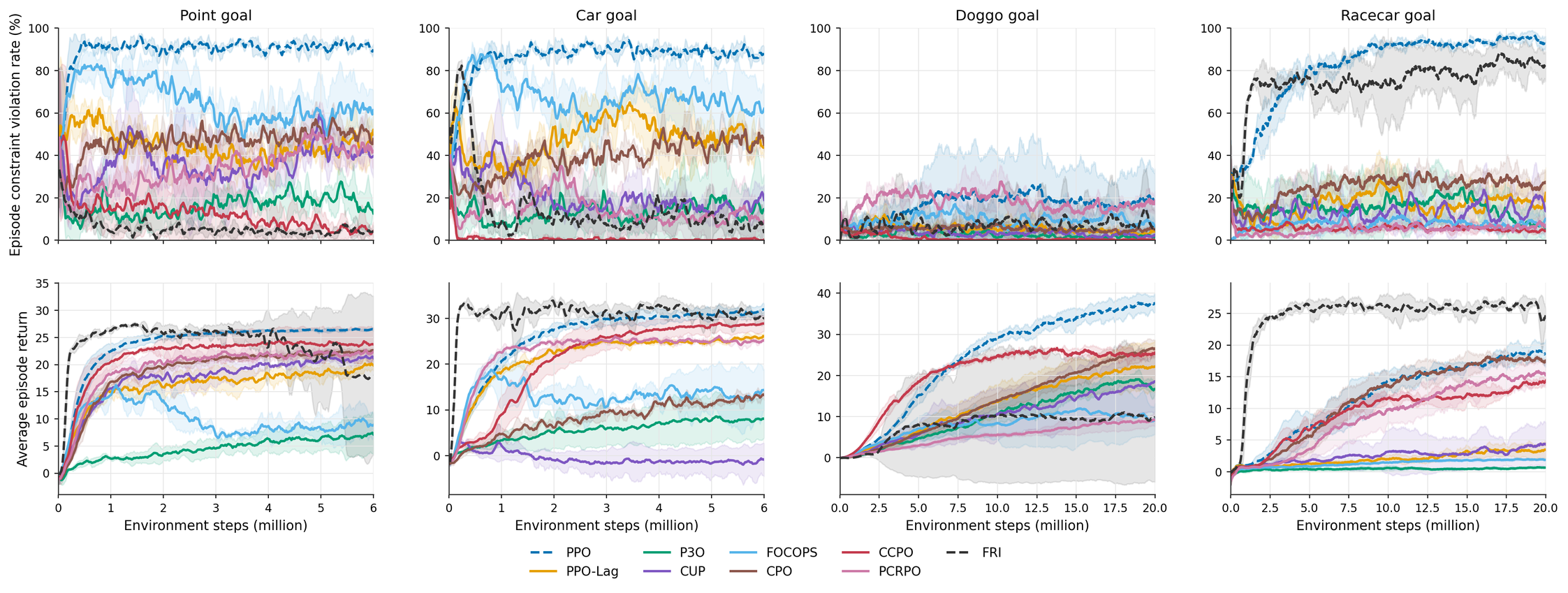}
	\caption{
		Training curves on four safety-critical control benchmarks. The shaded regions represent 95\% confidence intervals over five seeds. The first row shows the episode constraint violation rate(\%), and the second row shows the average episode return. The columns correspond to \emph{PointGoal}, \emph{CarGoal}, \emph{DoggoGoal}, and \emph{RacecarGoal}, respectively.
	}
	\label{fig:training_grid}
\end{figure*}

\subsection{Safety-Critical Control Benchmarks}
\label{sec:safety_control_benchmarks}

{We further evaluate CCPO on four safety-critical benchmarks with nonlinear dynamics and explicit safety constraints, namely \emph{PointGoal}, \emph{CarGoal}, \emph{DoggoGoal}, and \emph{RacecarGoal} {from Safety-Gymnasium~\cite{ji2023safety}}. The constrained baselines are \emph{CPO}~\cite{achiam2017cpo}, \emph{PPO-Lag}~\cite{ray2019benchmarking}, \emph{CUP}~\cite{2022NIPS_CUP}, \emph{FOCOPS}~\cite{zhang2020first}, and \emph{P3O}~\cite{zhang2022p3o}.
{We additionally include FRI~\cite{yang2023synthesizing}, using the original implementation based on soft actor--critic (SAC), an off-policy algorithm, and PCRPO~\cite{gu2024balance}, which addresses reward–safety gradient conflicts through gradient projection.}
Unconstrained \emph{PPO} is also included in Fig.~\ref{fig:training_grid} as a reference. All methods use comparable network architectures and training budgets, and we report the {episode constraint violation rate} and return over five random seeds.}

\blue{
\begin{table*}[t]
	\centering
	\caption{Average violation and return in the last 10\% of training iterations on safety-critical control benchmarks. Results are reported as mean $\pm$ standard deviation over five seeds. Bold denotes the best mean (lower violation, higher return).}
	\label{tab:benchmark_results}
	\resizebox{\textwidth}{!}{%
	\begin{tabular}{lcccccccc}
		\toprule
		& \multicolumn{2}{c}{PointGoal}
		& \multicolumn{2}{c}{CarGoal}
		& \multicolumn{2}{c}{DoggoGoal}
		& \multicolumn{2}{c}{RacecarGoal}\\
		\cmidrule(lr){2-3}\cmidrule(lr){4-5}\cmidrule(lr){6-7}\cmidrule(lr){8-9}
		Method & Viol. (\%) & Return & Viol. (\%) & Return & Viol. (\%) & Return & Viol. (\%) & Return\\
		\midrule
		CCPO & $4.91\pm3.67$ & $\bm{23.72\pm2.73}$ & $\bm{0.48\pm0.14}$ & $28.61\pm1.51$ & $\bm{0.30\pm0.23}$ & $25.06\pm0.93$ & $\bm{4.49\pm0.16}$ & $13.86\pm0.45$\\
		CPO & $50.17\pm0.40$ & $22.37\pm0.41$ & $46.89\pm2.88$ & $12.68\pm0.48$ & $4.79\pm0.75$ & $\bm{25.67\pm2.08}$ & $26.12\pm3.24$ & $17.70\pm0.32$\\
		PPO-Lag & $46.50\pm7.47$ & $19.25\pm0.85$ & $48.47\pm1.25$ & $25.75\pm0.47$ & $3.29\pm0.95$ & $21.91\pm5.05$ & $17.63\pm6.39$ & $3.28\pm1.04$\\
		FOCOPS & $61.82\pm10.02$ & $9.19\pm2.48$ & $62.44\pm12.94$ & $13.09\pm5.59$ & $7.83\pm2.90$ & $9.89\pm4.36$ & $8.06\pm3.00$ & $1.93\pm0.13$\\
		CUP & $41.11\pm3.47$ & $21.03\pm0.53$ & $16.50\pm4.47$ & $-1.36\pm3.13$ & $2.10\pm0.65$ & $17.51\pm6.18$ & $15.23\pm7.50$ & $4.01\pm2.80$\\
		P3O & $17.44\pm6.51$ & $6.78\pm3.18$ & $12.62\pm14.21$ & $7.79\pm4.02$ & $1.76\pm1.57$ & $18.16\pm1.21$ & $7.34\pm6.88$ & $0.62\pm0.10$\\
		PCRPO & $42.79\pm4.52$ & $22.03\pm0.14$ & $11.88\pm8.68$ & $24.87\pm0.48$ & $16.82\pm2.49$ & $8.78\pm1.50$ & $6.62\pm0.94$ & $15.24\pm1.62$\\
		\blue{FRI} & $\bm{4.78\pm1.56}$ & $18.09\pm12.86$ & $6.56\pm4.52$ & $\bm{30.71\pm0.88}$ & $8.29\pm10.06$ & $10.06\pm14.27$ & $83.55\pm6.31$ & $\bm{25.86\pm0.35}$\\
		\bottomrule
	\end{tabular}%
	}
\end{table*}
}
{As shown in Fig.~\ref{fig:training_grid}, CCPO maintains low episode constraint violation rates across all four tasks. On PointGoal and CarGoal, its violation rate decreases rapidly during early training and remains low thereafter. On DoggoGoal and RacecarGoal, it also achieves low violation rates toward the end of training.}

{Table~\ref{tab:benchmark_results} demonstrates that CCPO achieves a favorable balance between constraint satisfaction and task performance. It attains the lowest mean violation rate on three of the four benchmarks and remains close to the lowest rate on PointGoal, where it achieves the highest return among the compared methods. In particular, violation rates below 0.5\% on CarGoal and DoggoGoal are accompanied by returns close to the best baseline results, indicating that strong constraint satisfaction can be achieved without substantially compromising task performance.

Comparisons with {FRI} and PCRPO further clarify this trade-off. Despite the different learning paradigms---{FRI} is off-policy, whereas CCPO is on-policy---CCPO achieves a higher return on PointGoal at a similar violation rate and improves both metrics on DoggoGoal. On CarGoal, {FRI} attains a higher return, but CCPO reduces the violation rate from 6.56\% to 0.48\%, with a comparatively modest return decrease from 30.71 to 28.61. Relative to PCRPO, CCPO achieves both lower violation rates and higher returns on all three tasks. On RacecarGoal, CCPO instead favors constraint satisfaction, attaining a lower violation rate than either method at the expense of return.}

\section{Conclusion}
\label{sec:conclusion}

We introduced the Constraint Aggregation Function, a state- and policy-dependent representation of long-horizon safety with a discounted Bellman recursion and a conditional feasible-set interpretation. Its action-conditioned form yields a risk advantage that identifies actions whose predicted unsafety exceeds the current-policy baseline.
We proposed Cone-Constrained Policy Optimization. CCPO constructs a risk-descent direction from the CAF-induced risk advantage, resolves conflicts with task ascent through gradient projection, and selects the final update from the relative interior of the resulting safety cone.
We further established the monotonicity properties of the projected update. Under exact gradients, standard smoothness conditions, and a sufficiently small step size, the update increases task return while decreasing the policy-level CAF functional.
Numerical results show that CAF recovers the reference feasible boundary and that CCPO achieves feasible near-optimal control in emergency braking. Across four safety-critical benchmarks, CCPO maintains {low constraint violation rates} while preserving competitive returns relative to constrained baselines.

\bigskip
\appendices

\section{\texorpdfstring{{Proof of Proposition~\ref{prop:bellman_caf}}}{Proof of Bellman Consistency of CAF}}
\label{app:bellman_caf_proof}

\subsection{Proof of Proposition~\ref{prop:bellman_caf}}
Starting from the survival-weighted infinite series, we derive the one-step
identity consisting of the immediate term $s(x)$ and a discounted CAF tail from
the next state.
\begin{IEEEproof}
	Fix an arbitrary state $x\in\mathcal X$. Let the trajectory
	$\tau=(x_0,u_0,x_1,u_1,\ldots)$ be generated by
	$u_t\sim\pi(\cdot\mid x_t)$ and $x_{t+1}=f(x_t,u_t)$ with $x_0=x$.
	By the definition of CAF,
	\begin{align}
		F^\pi(x)
		=
		\mathbb E
		\left[
		\sum_{t=0}^{\infty}
		\gamma^t
		\left(
		\prod_{j=0}^{t-1}(1-s(x_j))
		\right)
		s(x_t)
		\,\middle|\, x_0=x
		\right].
	\end{align}
	The pathwise partial sums are nondecreasing and bounded by the telescoping
	survival-weight argument in Proposition~\ref{prop:caf_bounded}. Thus, the
	series is integrable, and monotone convergence justifies the following
	decomposition and conditional expectations.

	\emph{Step 1: Separate the immediate term from the tail.}
	We first separate the current risk from the future tail risk. The
	$t=0$ term equals $s(x)$, since $\gamma^0=1$, the empty product equals
	one, and $x_0=x$. For all $t\ge 1$, each tail term contains the common
	factor $\gamma(1-s(x))$. Define the tail sum
	\begin{equation}
		R_1
		\triangleq
		\sum_{t=1}^{\infty}
		\gamma^{t-1}
		\left(\prod_{j=1}^{t-1}(1-s(x_j))\right)
		s(x_t).
		\label{eq:app_tail_sum_r1}
	\end{equation}
	Hence,
	\begin{align}
		F^\pi(x)
		&=
		s(x)
		+
		\gamma(1-s(x))\,
		\mathbb E\!\left[R_1\,\middle|\, x_0=x\right].
	\end{align}
	Thus, CAF is decomposed into the immediate risk and the future risk
	propagated after the current state survives.

	\emph{Step 2: Identify the tail as a restarted CAF.}
	The remaining infinite sum $R_1$ has the same form as the CAF definition with
	$x_1$ as the new initial state:
	\begin{align}
		R_1
		&=
		\sum_{t=1}^{\infty}
		\gamma^{t-1}
		\left(
		\prod_{j=1}^{t-1}(1-s(x_j))
		\right)
		s(x_t)
		\notag\\
		&=
		s(x_1)
		+
		\gamma(1-s(x_1))s(x_2)
		+
		\cdots .
	\end{align}
	By the Markov property under the stationary policy $\pi$, conditioning on
	$x_1$ makes the future trajectory distributed as a fresh trajectory
	starting from $x_1$ under the same policy. Therefore,
	\begin{align}
		\mathbb E
		\left[
		R_1
		\,\middle|\, x_1
		\right]
		=
		F^\pi(x_1).
	\end{align}
	Using the tower property, we obtain
	\begin{align}
		\mathbb E
		\left[
		R_1
		\,\middle|\, x_0=x
		\right]
		=
		\mathbb E
		\left[
		F^\pi(x_1)
		\,\middle|\, x_0=x
		\right].
	\end{align}
	Substituting this into Step 1 gives
	\begin{align}
		F^\pi(x)
		=
		s(x)
		+
		\gamma(1-s(x))
		\mathbb E
		\left[
		F^\pi(x_1)
		\,\middle|\, x_0=x
		\right].
	\end{align}

	\emph{Step 3: Express the next-state expectation through the policy.}
	Since $x_1=f(x,u)$ under $x_0=x$ and $u\sim\pi(\cdot\mid x)$, the
	conditional expectation over $x_1$ can be written as an expectation over
	the action:
	\begin{align}
		\mathbb E
		\left[
		F^\pi(x_1)
		\,\middle|\, x_0=x
		\right]
		=
		\mathbb E_{u\sim\pi(\cdot\mid x)}
		\left[
		F^\pi(f(x,u))
		\right].
	\end{align}
	Therefore,
	\begin{align}
		F^\pi(x)
		=
		s(x)
		+
		\gamma(1-s(x))
		\mathbb E_{u\sim\pi(\cdot\mid x)}
		\left[
		F^\pi(f(x,u))
		\right],
	\end{align}
	which proves the Bellman-style recursion.
\end{IEEEproof}

\section{\texorpdfstring{{Proof of Theorem~\ref{thm:contraction_mapping}}}{Proof of the CAF Fixed-Point Theorem}}
\label{app:convergence_analysis_caf}

Under the conditions and notation of Theorem~\ref{thm:contraction_mapping}, the proof first establishes self-mapping and contraction. Uniqueness then identifies the fixed point with the series-defined CAF.
\begin{IEEEproof}
	\emph{Step 1: $\mathcal T^\pi$ maps the Banach space into itself.}
	For $F\in\mathcal B$, measurability of $s$, $f$, and the Markov kernel
	implies that $\mathcal T^\pi F$ is measurable. Moreover,
	\begin{equation}
		\|\mathcal T^\pi F\|_\infty
		\le 1+\gamma\|F\|_\infty<\infty,
	\end{equation}
	so $\mathcal T^\pi F\in\mathcal B$.

	\emph{Step 2: Contraction identifies the CAF fixed point.}
	For any $F,G\in\mathcal{B}$ and $x\in\mathcal{X}$,
	\begin{align}
	&\left|(\mathcal{T}^\pi F)(x)-(\mathcal{T}^\pi G)(x)\right|\notag\\
	&\quad\le
	\gamma(1-s(x))
	\mathbb{E}_{u}\!\left[
	\left|F(f(x,u))-G(f(x,u))\right|
	\right]\notag\\
	&\quad\le\gamma\|F-G\|_\infty.
	\end{align}
	Taking the supremum over $x$ proves the $\gamma$-contraction. Because
	$\mathcal B$ is complete, the Banach fixed-point theorem gives a unique fixed
	point and
	\begin{equation}
		\|F^{(k)}-F^\pi\|_\infty
		\le
		\gamma^k\|F^{(0)}-F^\pi\|_\infty.
	\end{equation}
	This holds for every initialization in $\mathcal B$. Under the stated
	measurability assumptions, finite-horizon CAF functions are measurable by
	backward recursion, and their pointwise monotone limit $F^\pi$ is measurable.
	Proposition~\ref{prop:caf_bounded} gives boundedness, so $F^\pi\in\mathcal B$.
	Proposition~\ref{prop:bellman_caf} then shows that $F^\pi$ is a fixed point of
	$\mathcal T^\pi$, and uniqueness identifies it with the Banach fixed point.
\end{IEEEproof}

\section{\texorpdfstring{{Proof of Proposition~\ref{prop:cone_acute_alignment}}}{Proof of Safety-Cone Acute Alignment}}
\label{app:cone_acute_alignment}

The proof has two steps. Step 1 shows that each rectified direction is orthogonal
to the other source direction and has strictly positive inner product with its own,
the latter being strict because $g_{\mathrm{task}}$ and $g_{\mathrm{risk}}$ are
non-collinear. Step 2 transfers both signs to an arbitrary
$g\in\operatorname{ri}(\mathcal C)$, where $\alpha>0$ and $\beta>0$.
\begin{IEEEproof}
	Fix $g\in\operatorname{ri}(\mathcal C)$, so
	\begin{equation}
		g=\alpha g_{\mathrm{task}}^++\beta g_{\mathrm{risk}}^+,
		\qquad \alpha>0,\ \beta>0.
		\label{eq:app_cone_decomp}
	\end{equation}

	\emph{Step 1: Establish positive self-alignment of both projected bases.}
	The projection definitions give the cross-orthogonality relations
	\begin{equation}
		\langle g_{\mathrm{task}}^+,g_{\mathrm{risk}}\rangle=0,
		\qquad
		\langle g_{\mathrm{risk}}^+,g_{\mathrm{task}}\rangle=0.
		\label{eq:app_projection_orthogonality}
	\end{equation}
	They also give
	\begin{align}
		\langle g_{\mathrm{task}}^+,g_{\mathrm{task}}\rangle
		&=\|g_{\mathrm{task}}\|_2^2
		-
		\frac{\langle g_{\mathrm{task}},g_{\mathrm{risk}}\rangle^2}
		{\|g_{\mathrm{risk}}\|_2^2}>0,
		\label{eq:app_task_basis_positive}\\
		\langle g_{\mathrm{risk}}^+,g_{\mathrm{risk}}\rangle
		&=\|g_{\mathrm{risk}}\|_2^2
		-
		\frac{\langle g_{\mathrm{task}},g_{\mathrm{risk}}\rangle^2}
		{\|g_{\mathrm{task}}\|_2^2}>0.
		\label{eq:app_risk_basis_positive}
	\end{align}
	The inequalities are strict because the original directions are nonzero and
	non-collinear, so Cauchy--Schwarz is strict.

	\emph{Step 2: Transfer positivity to every relative-interior direction.}
	Using \eqref{eq:app_cone_decomp} and the cross-orthogonality relations,
	\begin{align}
		\langle g,g_{\mathrm{task}}\rangle
		&=\alpha\langle g_{\mathrm{task}}^+,g_{\mathrm{task}}\rangle>0,
		\label{eq:app_task_alignment_positive}\\
		\langle g,g_{\mathrm{risk}}\rangle
		&=\beta\langle g_{\mathrm{risk}}^+,g_{\mathrm{risk}}\rangle>0.
		\label{eq:app_risk_alignment_positive}
	\end{align}
	Thus, $g$ forms acute angles with both the task-ascent direction and the
	risk-descent direction.
\end{IEEEproof}

\section{\texorpdfstring{{Proof of Theorem~\ref{thm:joint_monotonicity}}}{Proof of Simultaneous Task Improvement and Risk Reduction}}
\label{app:stepsize_bounds}

{Lemma~\ref{lem:descent_lemma} first decomposes the changes in the task return and \(\mathcal{R}\) into first-order terms and quadratic remainders.
We then convert these bounds into task-side and risk-side step-size conditions in Appendix~\ref{app:mono_task}; Appendix~\ref{app:complete_joint_monotonicity} combines them with cone alignment.}

\subsection{\texorpdfstring{{Admissible Step-Size Conditions for Theorem~\ref{thm:joint_monotonicity}}}{Admissible Step-Size Conditions for the Monotonicity Theorem}}
\label{app:mono_task}

We next convert the quadratic models into the task-side and risk-side
conditions needed by the theorem. The derivations are parallel because the same
step must increase task return and decrease $\mathcal R$.

Applying the lower model in Lemma~\ref{lem:descent_lemma} to
\(J_{\mathrm{task}}\) with \({\Delta}=\eta g\) gives
\begin{equation}
	J_{\mathrm{task}}(\theta_k+\eta g)-J_{\mathrm{task}}(\theta_k)
	\ge
	\eta\!\left(
	\langle g_{\mathrm{task}},g\rangle
	-\frac{L_{\mathrm{task}}}{2}\eta\|g\|_2^2
	\right).
	\label{eq:app_task_quadratic_change}
\end{equation}
{By \eqref{eq:task_risk_directions}, $\nabla\mathcal R(\theta_k)=-g_{\mathrm{risk}}(\theta_k)$.}
The upper model therefore yields
\begin{equation}
	\mathcal R(\theta_k+\eta g)
	-\mathcal R(\theta_k)
	\le
	\eta\!\left(
	-\langle g_{\mathrm{risk}},g\rangle
	+\frac{L_{\mathrm{risk}}}{2}\eta\|g\|_2^2
	\right).
	\label{eq:app_risk_quadratic_change}
\end{equation}
Suppose that \(g\) is positively aligned with the task-ascent and risk-descent directions.
Factoring out \(\eta>0\) in
\eqref{eq:app_task_quadratic_change}--\eqref{eq:app_risk_quadratic_change}
shows that strict task improvement and a strict decrease in $\mathcal R$ hold if
\begin{equation}
	0<\eta<
	\min\left\{
	\bar\eta_{\mathrm{task}}(g),
	\bar\eta_{\mathrm{risk}}(g)
	\right\},
	\label{eq:app_joint_step_condition}
\end{equation}
where the admissible upper bounds are those in
\eqref{eq:direct_step_bounds_main}:
\begin{equation}
	\bar\eta_{\mathrm{task}}(g)
	=
	\frac{2\langle g_{\mathrm{task}},g\rangle}
	{L_{\mathrm{task}}\|g\|_2^2},
	\qquad
	\bar\eta_{\mathrm{risk}}(g)
	=
	\frac{2\langle g_{\mathrm{risk}},g\rangle}
	{L_{\mathrm{risk}}\|g\|_2^2}.
	\label{eq:app_admissible_step_bounds}
\end{equation}

For a relative-interior cone direction
\(g=\alpha g_{\mathrm{task}}^++\beta g_{\mathrm{risk}}^+\),
cross-orthogonality makes the favorable first-order terms explicit:
\begin{equation}
	\langle g_{\mathrm{task}},g\rangle
	=
	\alpha\|g_{\mathrm{task}}^+\|_2^2,
	\qquad
	\langle g_{\mathrm{risk}},g\rangle
	=
	\beta\|g_{\mathrm{risk}}^+\|_2^2.
	\label{eq:app_cone_alignment_terms}
\end{equation}
Thus, Proposition~\ref{prop:cone_acute_alignment} supplies the positive
numerators required by \eqref{eq:app_admissible_step_bounds}. It remains only
to show that the coefficient choice in \eqref{eq:cone_coefficient_choice}
makes \eqref{eq:conservative_joint_stepsize} admissible for both the task-side and risk-side changes.

\subsection{\texorpdfstring{{Completion of the Proof of Theorem~\ref{thm:joint_monotonicity}}}{Completion of the Monotonicity Proof}}
\label{app:complete_joint_monotonicity}

We now combine the geometric and finite-step parts of the argument. Acute
alignment first establishes the signs of the two linear terms. The coefficient
choice then provides one task-side and one risk-side admissible upper bound, after which
the stated step size can be substituted into the quadratic models.
\begin{IEEEproof}
	\emph{Step 1: Establish positive alignment for the selected direction.}
	The coefficients in \eqref{eq:cone_coefficient_choice} are
	strictly positive because both source directions are nonzero. Hence,
	\(g_{\mathrm{cone}}\in\operatorname{ri}(\mathcal C)\), and
	Proposition~\ref{prop:cone_acute_alignment} gives
	\begin{equation}
		\langle g_{\mathrm{task}},g_{\mathrm{cone}}\rangle>0,
		\qquad
		\langle g_{\mathrm{risk}},g_{\mathrm{cone}}\rangle>0.
		\label{eq:app_selected_acute_alignment}
	\end{equation}
	The positive inner products also imply \(g_{\mathrm{cone}}\neq0\).
	Consequently, both upper bounds in
	\eqref{eq:app_admissible_step_bounds} have nonzero denominators and are
	strictly positive.

	\emph{Step 2: Verify the common step-size range.}
	Let
	\begin{equation}
		a\triangleq\|g_{\mathrm{task}}\|_2,
		\qquad
		b\triangleq\|g_{\mathrm{risk}}\|_2,
		\qquad
		\psi\triangleq
		\angle(g_{\mathrm{task}},g_{\mathrm{risk}}).
		\label{eq:app_abpsi_def}
	\end{equation}
	The conflict and non-collinearity assumptions give
	\(\psi\in(\pi/2,\pi)\), so \(\sin\psi>0\). The projection identities in
	\eqref{eq:projected_direction_norms} yield
	\begin{equation}
		\|g_{\mathrm{task}}^+\|_2=a\sin\psi,
		\qquad
		\|g_{\mathrm{risk}}^+\|_2=b\sin\psi.
		\label{eq:app_selected_projected_norms}
	\end{equation}
	For the coefficients in \eqref{eq:cone_coefficient_choice},
	\begin{equation}
		\alpha=\frac{b}{a+b},
		\qquad
		\beta=\frac{a}{a+b}.
		\label{eq:app_selected_coefficients}
	\end{equation}
	Substituting \eqref{eq:app_selected_projected_norms} and
	\eqref{eq:app_selected_coefficients} into
	\eqref{eq:app_cone_alignment_terms} gives the exact alignments
	\begin{equation}
		\langle g_{\mathrm{task}},g_{\mathrm{cone}}\rangle
		=
		\frac{ba^2\sin^2\psi}{a+b},
		\qquad
		\langle g_{\mathrm{risk}},g_{\mathrm{cone}}\rangle
		=
		\frac{ab^2\sin^2\psi}{a+b}.
		\label{eq:app_selected_alignments}
	\end{equation}
	The triangle inequality also gives
	\begin{align}
		\|g_{\mathrm{cone}}\|_2
		&\le
		\alpha\|g_{\mathrm{task}}^+\|_2
		+
		\beta\|g_{\mathrm{risk}}^+\|_2 \notag\\
		&=
		\frac{2ab\sin\psi}{a+b}.
		\label{eq:app_selected_direction_norm}
	\end{align}
	Combining \eqref{eq:app_admissible_step_bounds},
	\eqref{eq:app_selected_alignments}, and
	\eqref{eq:app_selected_direction_norm} yields
	\begin{align}
		\bar\eta_{\mathrm{task}}(g_{\mathrm{cone}})
		&\ge
		\frac{a+b}{2L_{\mathrm{task}}b}
		>
		\frac{1}{2L_{\mathrm{task}}},
		\label{eq:app_task_step_bound}\\
		\bar\eta_{\mathrm{risk}}(g_{\mathrm{cone}})
		&\ge
		\frac{a+b}{2L_{\mathrm{risk}}a}
		>
		\frac{1}{2L_{\mathrm{risk}}}.
		\label{eq:app_risk_step_bound}
	\end{align}
	The strict comparisons follow from \(a,b>0\). Thus, the coefficient choice
	makes the conservative range satisfy the task-side and risk-side requirements.

	\emph{Step 3: Apply the common step size to the task-return objective and $\blue{\mathcal R}$.}
	The condition \eqref{eq:conservative_joint_stepsize} and
	\eqref{eq:app_task_step_bound}--\eqref{eq:app_risk_step_bound}
	imply
	\begin{equation}
		0<\eta<
		\min\left\{
		\bar\eta_{\mathrm{task}}(g_{\mathrm{cone}}),
		\bar\eta_{\mathrm{risk}}(g_{\mathrm{cone}})
		\right\}.
		\label{eq:app_common_step_condition}
	\end{equation}
	Applying \eqref{eq:app_task_quadratic_change} and
	\eqref{eq:app_risk_quadratic_change} with \(g=g_{\mathrm{cone}}\) therefore
	gives
	\begin{align}
		J_{\mathrm{task}}(\theta_k+\eta g_{\mathrm{cone}})
		&>J_{\mathrm{task}}(\theta_k),\\
		{\mathcal R}(\theta_k+\eta g_{\mathrm{cone}})
		&<{\mathcal R}(\theta_k).
	\end{align}
	These are exactly the task-return increase and decrease of the {policy-level CAF criterion} claimed in
	Theorem~\ref{thm:joint_monotonicity}.
\end{IEEEproof}

{
\section{Proof of Finite-Step Improvement Guarantees}
\label{app:finite_step_convergence}
}
We give a concise {two-step} proof of {Theorem~\ref{thm:finite_step}}.
{Let $G_{\mathrm{risk}}\triangleq\max_{\Theta}\|g_{\mathrm{risk}}\|_2$, which, like $G_{\mathrm{task}}$, is finite by Assumption~\ref{assum:smoothness}, and let $G\triangleq G_{\mathrm{task}}+\max\{1,\lambda_{\mathrm{risk}}\}G_{\mathrm{risk}}$, which bounds $\|g_k\|_2$.}
{With $\bar L\triangleq\max\{1,\lambda_{\mathrm{risk}}\}(L_{\mathrm{task}}+L_{\mathrm{risk}})$, a sufficiently small step size means
\begin{equation}
\eta\le\min\{1/(4\bar L),\,C_\eta\epsilon_{\mathrm{risk}}^2\},
\label{eq:app_eta_conditions}
\end{equation}
where $C_\eta\triangleq\min\{\lambda_{\mathrm{risk}}c_{\mathrm{risk}}^2/(4L_{\mathrm{risk}}G^2),\,4/(4G_{\mathrm{task}}c_{\mathrm{risk}}+G^2)\}$ depends only on the problem constants.
Let $a$, $b$, and $\psi$ be as in \eqref{eq:app_abpsi_def} at $\theta_k$. By \eqref{eq:app_selected_alignments} and \eqref{eq:app_selected_direction_norm}, $\|g_{\mathrm{cone}}\|_2^2$ is at most $4\langle g_{\mathrm{task}},g_{\mathrm{cone}}\rangle$ and at most $4\langle g_{\mathrm{risk}},g_{\mathrm{cone}}\rangle$. Lemma~\ref{lem:descent_lemma} and $\eta\le1/(4\bar L)$ then give
\begin{equation}
\begin{aligned}
J_{\mathrm{task}}(\theta_{k+1})-J_{\mathrm{task}}(\theta_k)&\ge\tfrac{\eta}{2}\langle g_{\mathrm{task}},g_k\rangle,\\
\mathcal R(\theta_k)-\mathcal R(\theta_{k+1})&\ge\tfrac{\eta}{2}\langle g_{\mathrm{risk}},g_k\rangle
\end{aligned}
\label{eq:app_sufficient}
\end{equation}
for $g_k=g_{\mathrm{cone}}$. The first bound also holds for $g_k=g_{\mathrm{task}}$, and the second for $g_k=g_{\mathrm{risk}}$.}

\emph{{Step 1: Establish finite risk-region entry and retention.}}
{Let $\delta_k\triangleq\mathcal R(\theta_k)-\mathcal R^\star_\Theta$, which lies in $[0,1]$ by Proposition~\ref{prop:caf_bounded}, and let $K_{\mathrm{in}}\triangleq\inf\{k:\delta_k\le\epsilon_{\mathrm{risk}}/2\}$.}
For every $k<K_{\mathrm{in}}$,
\eqref{eq:risk_gradient_domination} gives
{$b\ge c_{\mathrm{risk}}\delta_k>c_{\mathrm{risk}}\epsilon_{\mathrm{risk}}/2$.}
{Since $\varepsilon_{\mathrm{proj}}<c_{\mathrm{risk}}\epsilon_{\mathrm{risk}}/2$, it follows that $b>\varepsilon_{\mathrm{proj}}$, so the no-update and task-only branches are not taken, and the remaining branches give $\langle g_{\mathrm{risk}},g_k\rangle\ge c_0b^2$ with $c_0\triangleq\min\{\lambda_{\mathrm{risk}},\varepsilon_{\mathrm{proj}}\varepsilon_{\mathrm{ang}}^2/(2G_{\mathrm{risk}})\}$, where the last term arises from the cone update and uses $\varepsilon_{\mathrm{proj}}<b\le G_{\mathrm{risk}}$. In the risk-only and cone branches, \eqref{eq:app_sufficient} then gives $\delta_k-\delta_{k+1}\ge\frac{\eta}{2}c_0b^2$. In the non-conflict branch, Lemma~\ref{lem:descent_lemma}, $\|g_k\|_2\le G$, $b>c_{\mathrm{risk}}\epsilon_{\mathrm{risk}}/2$, and \eqref{eq:app_eta_conditions} give $\delta_k-\delta_{k+1}\ge\eta\lambda_{\mathrm{risk}}b^2-\frac{L_{\mathrm{risk}}}{2}\eta^2G^2\ge\frac{\eta}{2}\lambda_{\mathrm{risk}}b^2$. With $q\triangleq\eta c_0c_{\mathrm{risk}}^2/2$ and $b\ge c_{\mathrm{risk}}\delta_k$, it follows that}
\begin{equation}
{\delta_{k+1}\le\delta_k-q\delta_k^2,
\qquad
k<K_{\mathrm{in}}.}
\label{eq:finite_risk_decrease}
\end{equation}
{Hence $1/\delta_{k+1}\ge1/\delta_k+q$ whenever $\delta_{k+1}>0$, and $\delta_0\le1$ gives}
\begin{equation}
K_{\mathrm{in}}
\le
\left\lceil
{\frac{4}{\eta c_0c_{\mathrm{risk}}^2\epsilon_{\mathrm{risk}}}}
\right\rceil .
\label{eq:finite_entry_bound}
\end{equation}
{Retention follows by induction. If $\delta_k>\epsilon_{\mathrm{risk}}/2$, the argument above gives $\delta_{k+1}<\delta_k$. Otherwise, by \eqref{eq:app_sufficient}, only the task-only and non-conflict branches can increase $\mathcal R$, and Lemma~\ref{lem:descent_lemma}, \eqref{eq:app_eta_conditions}, and $\varepsilon_{\mathrm{proj}}<c_{\mathrm{risk}}\epsilon_{\mathrm{risk}}/2$ bound the increase by $\eta G_{\mathrm{task}}\varepsilon_{\mathrm{proj}}+\frac{L_{\mathrm{risk}}}{2}\eta^2G^2\le\epsilon_{\mathrm{risk}}/2$. Hence}
\begin{equation}
\mathcal R(\theta_k)
\le
{\mathcal R^\star_\Theta}+\epsilon_{\mathrm{risk}},
\qquad
k\ge K_{\mathrm{in}}.
\label{eq:finite_risk_retention_proof}
\end{equation}

\emph{{Step 2: Establish finite first-order stationarity.}}
For $k\ge K_{\mathrm{in}}$, let
\[
S_k
\triangleq
\min_{\mu\ge0}
\left\|
g_{\mathrm{task}}(\theta_k)
+\mu g_{\mathrm{risk}}(\theta_k)
\right\|_2 .
\]
{Whenever $S_k>\epsilon_{\mathrm{task}}$, we have $\|g_{\mathrm{task}}\|_2\ge S_k>\varepsilon_{\mathrm{proj}}$ and, under conflict (where $S_k=\|g_{\mathrm{task}}^+\|_2$), $\sin\psi=S_k/\|g_{\mathrm{task}}\|_2>\varepsilon_{\mathrm{ang}}$. Hence neither the no-update branch nor a risk-only branch is taken. Let $\zeta\triangleq\min\{\lambda_{\mathrm{risk}},1/2\}$ and $\Phi\triangleq J_{\mathrm{task}}-\zeta\mathcal R$, whose gradient $g_{\mathrm{task}}+\zeta g_{\mathrm{risk}}$ is $(L_{\mathrm{task}}+\zeta L_{\mathrm{risk}})$-Lipschitz. In the task-only branch, $b\le\varepsilon_{\mathrm{proj}}<a$ gives $\langle\nabla\Phi,g_k\rangle\ge a^2-\zeta ab\ge a^2/2=\|g_k\|_2^2/2$. In the non-conflict branch, $\langle\nabla\Phi,g_k\rangle\ge a^2$ and $\|g_k\|_2^2\le\max\{1,\lambda_{\mathrm{risk}}/\zeta\}\langle\nabla\Phi,g_k\rangle$. As $\max\{2,\lambda_{\mathrm{risk}}/\zeta\}(L_{\mathrm{task}}+\zeta L_{\mathrm{risk}})\le2\bar L$, Lemma~\ref{lem:descent_lemma} gives $\Phi(\theta_{k+1})-\Phi(\theta_k)\ge\frac{\eta}{2}\langle\nabla\Phi,g_k\rangle\ge\frac{\eta}{4}S_k^2$ in both branches. In the cone branch, \eqref{eq:app_sufficient} and \eqref{eq:app_selected_alignments} give $\Phi(\theta_{k+1})-\Phi(\theta_k)\ge\frac{\eta}{2}\frac{b}{a+b}S_k^2$, since $S_k=a\sin\psi$. As $b>\varepsilon_{\mathrm{proj}}$ in this branch and $\varepsilon_{\mathrm{proj}}<a\le G_{\mathrm{task}}$, so that $b/(a+b)>\varepsilon_{\mathrm{proj}}/(2G_{\mathrm{task}})$, every branch yields}
\begin{equation}
{\Phi(\theta_{k+1})-\Phi(\theta_k)
\ge
\frac{\eta\varepsilon_{\mathrm{proj}}\epsilon_{\mathrm{task}}^2}{4G_{\mathrm{task}}}.}
\label{eq:finite_task_progress}
\end{equation}
Since {$\Phi$ varies by at most $D\triangleq2r_{\max}/(1-\gamma)+1$ on $\Theta$},
\eqref{eq:finite_task_progress} cannot hold indefinitely.
Thus there exists a finite $k^\star\ge K_{\mathrm{in}}$ such that
$S_{k^\star}\le\epsilon_{\mathrm{task}}$, with
\begin{equation}
k^\star-K_{\mathrm{in}}
\le
\left\lceil
{\frac{4DG_{\mathrm{task}}}{\eta\varepsilon_{\mathrm{proj}}\epsilon_{\mathrm{task}}^2}}
\right\rceil .
\label{eq:finite_task_bound}
\end{equation}
Combining \eqref{eq:finite_entry_bound} and
\eqref{eq:finite_task_bound} yields the total iteration bound
\begin{equation}
k^\star
\le
\left\lceil
{\frac{4}{\eta c_0c_{\mathrm{risk}}^2\epsilon_{\mathrm{risk}}}}
\right\rceil
+
\left\lceil
{\frac{4DG_{\mathrm{task}}}{\eta\varepsilon_{\mathrm{proj}}\epsilon_{\mathrm{task}}^2}}
\right\rceil .
\label{eq:finite_step_bound}
\end{equation}


\end{document}